# An open boundary condition for high-order solutions of magnetohydrodynamics on unstructured grids

Xiaoliang Zhang (zxlnwpu@gmail.com) and Chunlei Liang (cliang@clarkson.edu)

*Mechanical and Aeronautical Engineering, Clarkson University, Potsdam, NY*

**In this paper a characteristics-based open boundary condition (CBC) is proposed for the magnetohydrodynamic (MHD) system of equations. The algorithm is carefully designed and implemented in the context of a high-order flux reconstruction (FR) scheme under the Generalized Lagrange Multiplier (GLM)-MHD system of equations. It is implemented by adding the contribution of the characteristic equation directly to the corrected flux term in the FR scheme dispensing with solving time- dependent characteristic equations along boundary faces. The CBC method is shown to be more accurate and robust than commonly used zero normal derivative (ZND) and approximate Riemann solver boundary conditions (ARBC) in solving 1D, 2D, and 3D test problems. The CBC method is successfully applied to simulate challenging problems of magnetic reconnection for which other options failed to get stable results over long-period time integration.**

## I. Introduction

It is often very difficult to represent the entire physical domain in modeling magnetic reconnection, the global dynamo phenomenon in the solar convection zone, and simulating plasma propulsion. Therefore, users and developers of MHD solvers would investigate artificial and "open" boundaries to reduce the size of the computational domain and allocate more computational elements to regions of interest. A highly relevant publication to our present work is the design of lacuna-based open boundary conditions by Meier et al.[1] However, in order to explicitly clean the divergence error of a magnetic field, this work has employed different mathematical models[2] from that of Meier et al. For all calculations presented in this paper, high-order flux reconstruction (FR) schemes are used on unstructured grids with all hexahedral elements to solve such MHD equations.[2, 3] Importantly, this paper proposes a new type of open boundary condition for the MHD system of equations.

## II. MHD Equations

In this research, a Generalized Lagrange Multiplier (GLM)-MHD formulation originally proposed by Dedner et al is employed.[4] In the GLM-MHD system, one auxiliary transport equation is appended to the original ideal MHD system of equations. The propagation of divergence error of magnetic field behaves similarly to propagating waves of hyperbolic equations. The GLM-MHD system in 3D physical space is shown below

$$\frac{\partial \rho}{\partial t} + \nabla \cdot (\rho \boldsymbol{u}) = 0, \tag{1}$$

$$\frac{\partial \rho \boldsymbol{u}}{\partial t} + \nabla \cdot \left(\rho \boldsymbol{u}\boldsymbol{u}^T + p\boldsymbol{I} + \frac{1}{2\mu}\boldsymbol{B}^2\boldsymbol{I} - \frac{1}{\mu}\boldsymbol{B}\boldsymbol{B}^T\right) = 0, \tag{2}$$

$$\frac{\partial \boldsymbol{B}}{\partial t} + \nabla \cdot \left(\boldsymbol{u}\boldsymbol{B}^T - \boldsymbol{B}\boldsymbol{u}^T + \phi\boldsymbol{I} - \eta_c(\nabla\boldsymbol{B} - \nabla\boldsymbol{B}^T)\right) = 0, \tag{3}$$

$$\frac{\partial \rho E}{\partial t} + \nabla \cdot \left(\boldsymbol{u} \cdot \left(\rho E + p + \frac{1}{2\mu}\boldsymbol{B}^2\right)\boldsymbol{I} - \frac{1}{\mu}\boldsymbol{B}\boldsymbol{B}^T - \frac{\eta_c}{\mu}\boldsymbol{B} \cdot (\nabla\boldsymbol{B} - \nabla\boldsymbol{B}^T)\right) = 0, \tag{4}$$

$$\frac{\partial \phi}{\partial t} + \nabla \cdot (c_h^2 \boldsymbol{B}) = 0, \tag{5}$$

where $\rho$ is density, $\mathbf{u}$ is the velocity field, $\mathbf{B}$ is the magnetic field, $p$ is pressure, $\mu$ is magnetic permeability, $\eta_c$ is conductivity, $E$ is the total energy, $e$ is the internal energy, $\varphi$ is the auxiliary variable introduced to form artificial divergence error wave, and $c_h$ is the wave speed.

The above system of equations is closed with the ideal gas law $p = \rho RT$ where $R$ is the gas constant.

In the Cartesian coordinates system, the above coupled nine time-dependent equations can be further written in a simple vectorial form,

$$\frac{\partial \boldsymbol{Q}}{\partial t} + \frac{\partial \boldsymbol{F}}{\partial x} + \frac{\partial \boldsymbol{G}}{\partial y} + \frac{\partial \boldsymbol{H}}{\partial z} = 0. \tag{6}$$

where $\mathbf{Q}$ is unknown vector, $\mathbf{Q} = [\rho, u, v, w, B_x, B_y, B_z, \rho E, \varphi]^T$.

# III. Flux Reconstruction (FR) Numerical Method

In this research, the FR method is employed to discretize the aforementioned GLM-MHD system of equations on unstructured hexahedral meshes. The FR method was originally introduced by Huynh[3] to solve conservation laws in nodal differential forms. It can flexibly be recovered to several popular high-order schemes, e.g., spectral difference (SD) and discontinuous Galerkin (DG) methods, by simply choosing corresponding correction functions. Moreover, the FR method is more computationally efficient than the SD or DG methods due to the following facts: (a) it uses coinciding interior flux and solution points (FPs and SPs) instead of two different sets of FPs and SPs involved in SD; (b) it is derived based on the differential form of governing equations with no need for calculating numerical integrals involved in DG.

## III.A. Coordinates Transformation

A twenty-node cubic iso-parametric mapping shown in Figure 1 is employed to transfer any physical position within each cell in the physical domain, $(x,y,z,t)$, to a reference position within a standard computational element in the computational domain, $(\xi, \eta, \zeta) \in [0, 1]\times[0, 1]\times[0, 1]$, $\tau = t$. Such a mapping procedure allows universal polynomial reconstructions regardless of the actual size/shape of the physical domain of interest. The transformation can be described as,

$$\begin{pmatrix} x \\ y \\ z \end{pmatrix} = \sum_{i=1}^{K} M_i\,(\xi, \eta, \zeta) \begin{pmatrix} x_i \\ y_i \\ z_i \end{pmatrix}, \tag{7}$$

where $K$ is the number of nodes defining a physical element, and $M_i$ are shape functions. The corresponding Jacobian matrix and its inverse are

$$\begin{aligned} \mathcal{J} &= \frac{\partial(x,y,z)}{\partial(\xi,\eta,\zeta)} = \begin{bmatrix} x_\xi & x_\eta & x_\zeta \\ y_\xi & y_\eta & y_\zeta \\ z_\xi & z_\eta & z_\zeta \end{bmatrix}, \\ \mathcal{J}^{-1} &= \frac{\partial(\xi,\eta,\zeta)}{\partial(x,y,z)} = \begin{bmatrix} \xi_x & \xi_y & \xi_z \\ \eta_x & \eta_y & \eta_z \\ \zeta_x & \zeta_y & \zeta_z \end{bmatrix} = \frac{1}{|\mathcal{J}|}\mathcal{S}, \end{aligned} \tag{8}$$

where S is the transpose of the cofactor matrix with vectorial components of $\vec{S}_\xi = |\mathbf{J}|[\xi_x, \xi_y, \xi_t]$, $\vec{S}_\eta = |\mathbf{J}|[\eta_x, \eta_y, \eta_t]$, $\vec{S}_\tau = |\mathbf{J}|[0, 0, 1]$, and $|\mathbf{J}|$ is the determinant of $\mathbf{J}$.

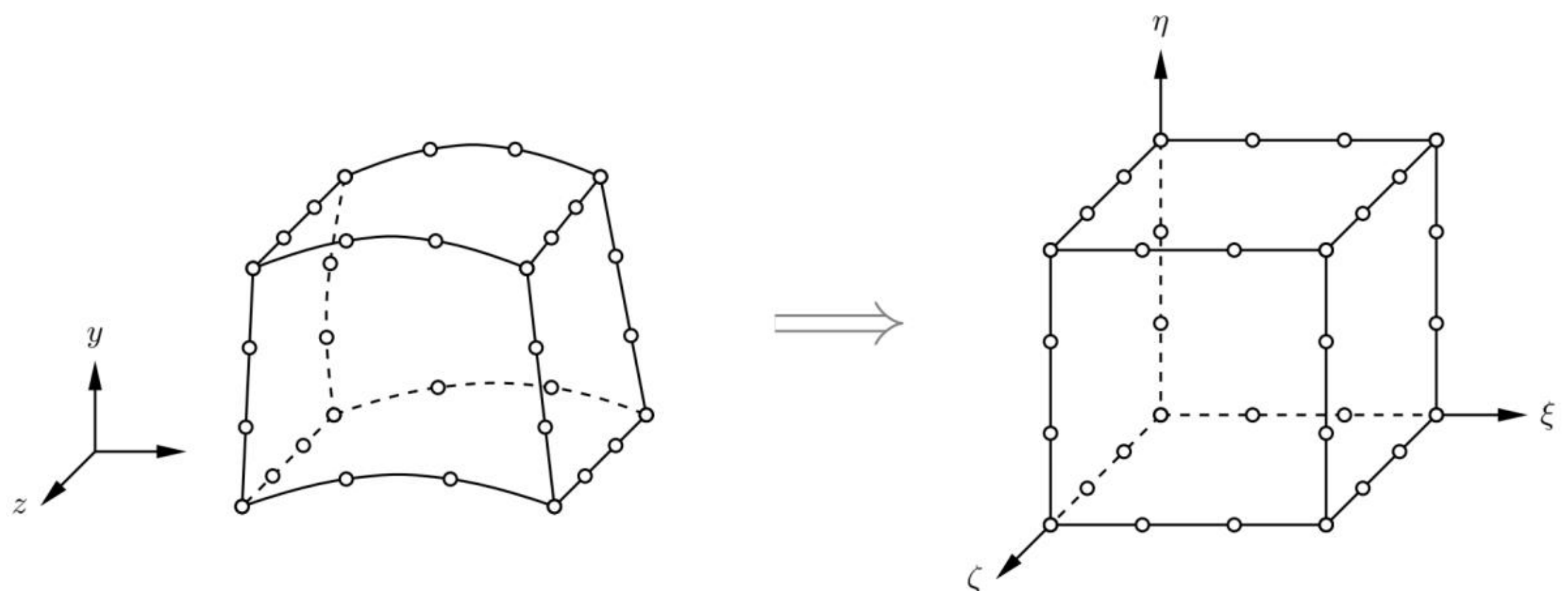

**Figure 1. A cubic iso-parametric mapping from a physical element to a standard computational cube**

## III.B. The Governing Equations in Computational Domain

Because of the mapping, the nine equations for the physical domain shown in Equation (6) also need to be transformed into a computational form. The transformed equations for the computational domain take the form

$$\frac{\partial \widetilde{\boldsymbol{Q}}}{\partial t}+\frac{\partial \widetilde{\boldsymbol{F}}}{\partial \xi}+\frac{\partial \widetilde{\boldsymbol{G}}}{\partial \eta}+\frac{\partial \widetilde{\boldsymbol{H}}}{\partial \zeta}=0, \tag{9}$$

where

$$\begin{pmatrix} \widetilde{\mathbf{F}} \\ \widetilde{\boldsymbol{G}} \\ \widetilde{\boldsymbol{H}} \end{pmatrix} = |\mathcal{J}|\mathcal{J}^{-1}\begin{pmatrix} \boldsymbol{F} \\ \boldsymbol{G} \\ \boldsymbol{H} \end{pmatrix}, \tag{10}$$

$$\widetilde{\boldsymbol{Q}} = |\mathcal{J}|\boldsymbol{Q}. \tag{11}$$

In this research, stationary meshes are only considered, which means that the Jacobian matrix is time independent. After applying the chain rule to the right-hand-side of Equation (11), one can obtain

$$\frac{\partial\, \widetilde{\boldsymbol{Q}}}{\partial \tau}=\frac{\partial(|\mathcal{J}|\boldsymbol{Q})}{\partial \tau}=|\mathcal{J}|\frac{\partial \boldsymbol{Q}}{\partial \tau}+\boldsymbol{Q}\frac{\partial|\mathcal{J}|}{\partial \tau}=|\mathcal{J}|\frac{\partial \boldsymbol{Q}}{\partial \tau}. \tag{12}$$

After substituting Equation (12) into Equation (9) and rearranging, one can obtain a new governing equation that was eventually implemented in the code with the following mathematical form,

$$\frac{\partial \boldsymbol{Q}}{\partial \tau} \quad = \frac{-1}{|\mathcal{J}|}\left(\frac{\partial\, \widetilde{\boldsymbol{F}}}{\partial \xi}+\frac{\partial\, \widetilde{\boldsymbol{G}}}{\partial \eta}+\frac{\partial\, \widetilde{\boldsymbol{H}}}{\partial \zeta}\right)=\mathscr{R}, \tag{13}$$

where $\mathscr{R}$ denotes the residual term.

## III.C. Spatial Discretization

In the FR method, both solution and flux polynomials are constructed through discrete and coincided points, i.e., SPs and FPs, using Lagrange interpolating basis functions. The SPs are placed on the roots of Legendre-Gauss polynomials, which are distributed inside a computational cell. $N$ quadrature points are used to construct ($N$-1)th order polynomials in each direction through Lagrange interpolating basis functions,

$$h_i(x) = \prod_{s=1,s\neq i}^{N} \left(\frac{X-X_s}{X_i-X_s}\right). \tag{14}$$

The reconstructed polynomial for unknown variables in a standard 3D computational element is a tensor product of three 1D polynomials,

$$\boldsymbol{Q}(\xi,\eta,\zeta) = \sum_{k=1}^{N}\sum_{j=1}^{N}\sum_{i=1}^{N} \boldsymbol{Q}_{i,j,k} h_i(\xi)\cdot h_j(\eta)\cdot h_k(\zeta). \tag{15}$$

The reconstructed flux polynomials are a series of element-wise continuous functions,

$$\begin{aligned}
\widetilde{\boldsymbol{F}}(\xi,\eta,\zeta) &= \sum_{k=1}^{N}\sum_{j=1}^{N}\sum_{i=1}^{N} \widetilde{\boldsymbol{F}}_{i,j,k} h_i(\xi)\cdot h_j(\eta)\cdot h_k(\zeta),\\
\widetilde{\boldsymbol{G}}(\xi,\eta,\zeta) &= \sum_{k=1}^{N}\sum_{j=1}^{N}\sum_{i=1}^{N} \widetilde{\boldsymbol{G}}_{i,j,k} h_i(\xi)\cdot h_i(\eta)\cdot h_k(\zeta),\\
\widetilde{\boldsymbol{H}}(\xi,\eta,\zeta) &= \sum_{k=1}^{N}\sum_{j=1}^{N}\sum_{i=1}^{N} \widetilde{\boldsymbol{H}}_{i,j,k} h_i(\xi)\cdot h_i(\eta)\cdot h_k(\zeta).
\end{aligned} \tag{16}$$

Figure 2 is a schematic of the distributions of SPs and FPs for the fourth-order FR scheme in 2D. **F**, **G**, **H** are stored on the same set of quadrature points.

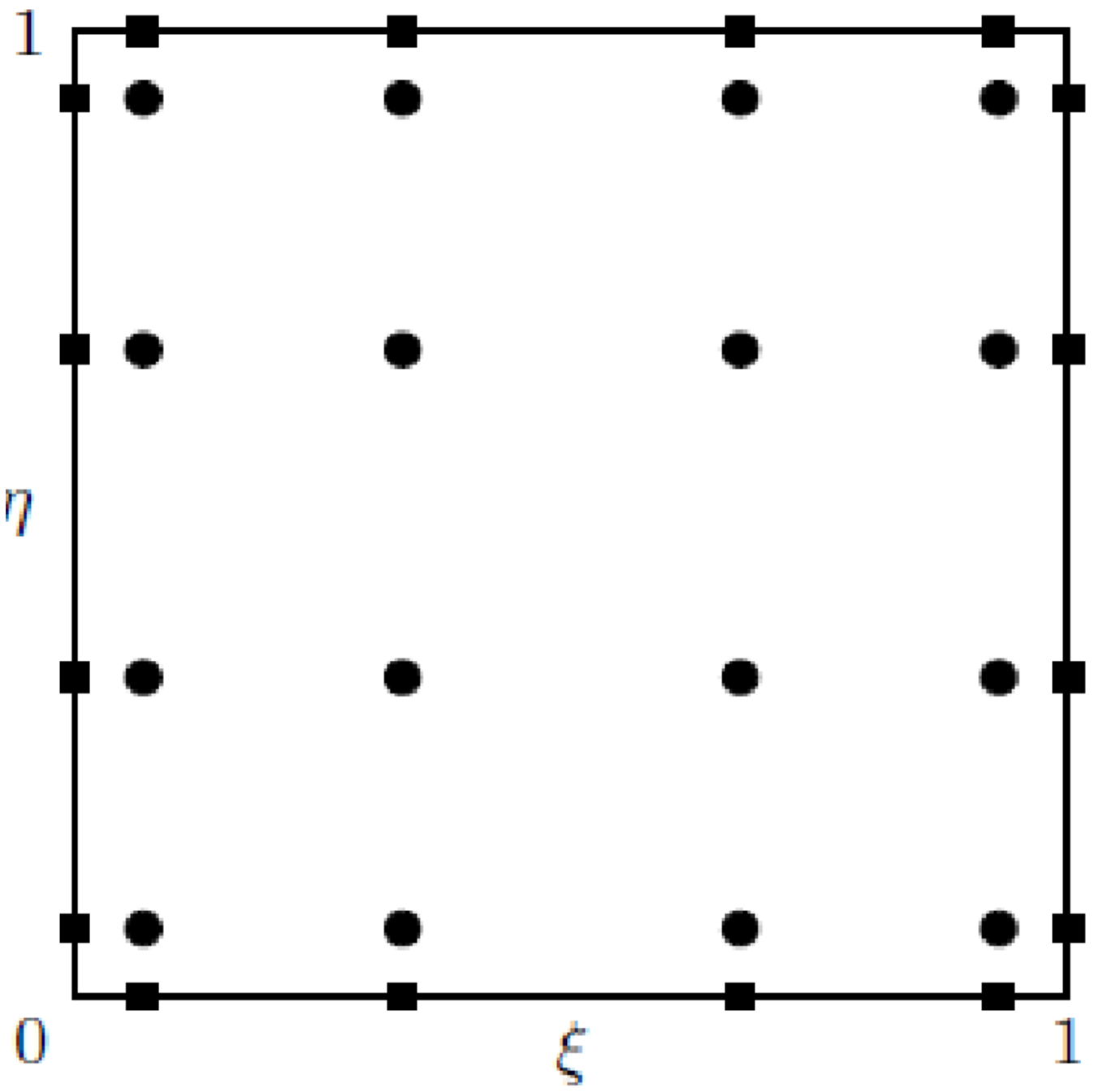

**Figure 2. A schematic of the distributions of solution points (SPs) and flux points (FPs) for the 4th order FR scheme where SPs are shown in circles and FPs are drawn in squares.**

The FPs located along cell interfaces are responsible for calculating the common fluxes on the interfaces of two adjacent cells. An approximate Riemann solver can be utilized for this purpose. In this paper, common fluxes are calculated through the Rusanov solver (also known as the local Lax-Friedrichs solver),[5]

$$\widetilde{\boldsymbol{F}}_{inv}^{com} = \frac{1}{2}\{\boldsymbol{F}_{inv}^{L} + \boldsymbol{F}_{inv}^{R} - (|\overline{\boldsymbol{V}}_n| + \bar{c}) \cdot (\boldsymbol{Q}_R - \boldsymbol{Q}_L) \cdot |\overrightarrow{S_\xi}| \cdot sign(\vec{n} \cdot \overrightarrow{S_\xi})\}, \quad (17)$$

where $\vec{n}$ is the normal direction of interfaces, $\overline{\mathbf{V}}_n$ is the normal component of velocity, and $\bar{c}$ is the speed of sound. $\widetilde{\boldsymbol{G}}_{inv}^{com}$ and $\widetilde{\boldsymbol{H}}_{inv}^{com}$ can also be obtained using the similar formulation for 3D calculations. The common values of unknown variables along an interface are the arithmetic average of solutions from the left and the right elements (also known as the BR1 scheme[6])

$$\boldsymbol{Q}^{com} = \frac{1}{2}(\boldsymbol{Q}^{L} + \boldsymbol{Q}^{R}) \quad (18)$$

The common viscous fluxes can be computed from the common solution values and common gradient values

$$\widetilde{\boldsymbol{F}}_{vis}^{com} = \widetilde{\boldsymbol{F}}_{vis}(\boldsymbol{Q}^{com}, \nabla\boldsymbol{Q}^{com}), \quad (19)$$

where $\nabla\boldsymbol{Q}^{com} = \nabla\boldsymbol{Q}^{L} + \nabla\boldsymbol{Q}^{R}$. Alternatively, the common viscous fluxes can be also computed via arithmetic average of left and right values,

$$\widetilde{\boldsymbol{F}}_{vis}^{com} = \frac{1}{2}(\widetilde{\boldsymbol{F}}_{vis}^{L} + \widetilde{\boldsymbol{F}}_{vis}^{R}). \quad (20)$$

A correction procedure is then applied to reconstruct a continuous flux polynomial. For instance, the continuous flux polynomial in $\xi$ direction can be mathematically described as,

$$\widetilde{\boldsymbol{F}}_{j,k}^{con}(\xi) = \widetilde{\boldsymbol{F}}_{j,k}(\xi) + \left[\widetilde{\boldsymbol{F}}_{j,k}^{com,L} - \widetilde{\boldsymbol{F}}_{j,k}(0)\right]g_{LB}(\xi) + \left[\widetilde{\boldsymbol{F}}_{j,k}^{com,R} - \widetilde{\boldsymbol{F}}_{j,k}(1)\right]g_{RB}(\xi) = \widetilde{\boldsymbol{F}}_{j,k}(\xi) + \widetilde{\boldsymbol{F}}_{j}^{c}(\xi), \quad (21)$$

where superscript 'con', 'c' denote 'continuous' and 'corrected', respectively, $\widetilde{\boldsymbol{F}}_{j,k}^{com,L}$ and $\widetilde{\boldsymbol{F}}_{j,k}^{com,R}$ are common flux on the left and right side of one cell interface, $\mathbf{F}_{j,k}(\xi)$ is an element-wise continuous flux approximation, and $j$, $k$ are directional indices along $\eta$ and $\zeta$ directions, respectively.

The correction function $g$ of interest is constructed using the same form as the $g_{DG}$ function in Huynh.[7] For example, the function on the left side of one cell interface is defined as,

$$g_{DG,LB} = \frac{(-1)^N}{2}(P_N - P_{N-1}), \quad (22)$$

where $P_N$ represents the n-th order Legendre polynomial. $g_{DG,LB}$ is required to be 1 at $\xi = 0$ and 0 at $\xi = 1$. The values of the n-th order $g_{DG}$ are required to be zeros at N-1 Legendre-Gauss quadrature points. These points are roots of $P_N-1$. These points normally have different locations from the roots of $P_N$, which are the locations of solution points of the N-th order. In other words, $g_{DG}$ requires correction at all interior solution points. The same procedures can be taken to obtain the continuous flux of $\widetilde{\boldsymbol{G}}_{i,k}^{con}(\eta)$ and $\widetilde{\boldsymbol{H}}_{i,j}^{con}(\zeta)$

# IV. Open Boundary Conditions

Three different open boundaries, ZND, ARBC, and CBC, are carefully designed and implemented in the context of the flux reconstruction (FR) scheme for solving the GLM-MHD system of equations. All three open boundaries are directly applied to the corrected flux term in the FR method. This process is computationally efficient especially for the CBC method since there is no need to solve time-dependent characteristic equations along boundary faces any more. The performance of three boundary condition candidates are evaluated and compared thoroughly via one, two and three-dimensional test cases in terms of their accuracy and robustness.

## IV.A. ZND formulation

The ZND boundary condition assumes zero gradient for all unknown variables on the boundary faces. Its mathematical formulation is shown in Equation (23).

$$\frac{\partial \phi}{\partial n} = 0, \tag{23}$$

where $\phi$ is a general dependent unknown variable.

It is widely used in numerical solvers due to its simplicity of implementation. However, it can generate significant amounts of reflections from the boundary. This method is a good reference for performing comparison studies against more accurate open boundary methods.

### IV.B. ARBC formulation

In the ARBC method, only the normal flux is specified on open boundary faces. Therefore, only 1D governing equation is considered which can be written as

$$\frac{\partial \boldsymbol{Q}}{\partial t} + \frac{\partial \boldsymbol{F}_n}{\partial n} = 0, \tag{24}$$

The residual term $\frac{\partial \boldsymbol{F}_n}{\partial n}$ can be further expanded using the chain rule as follows,

$$\frac{\partial \boldsymbol{F}_n}{\partial n} = \mathbb{R}\boldsymbol{\Lambda}\mathbb{L}\frac{\partial \boldsymbol{Q}}{\partial n} \tag{25}$$

where $\mathbb{R}\boldsymbol{\Lambda}\mathbb{L}$ is a product form of the flux Jacobian matrix. $\mathbb{R}$ and $\mathbb{L}$ are matrices of right and left eigenvectors, respectively and $\boldsymbol{\Lambda}$ is the diagonal matrix of eigenvalues.

In this research, the Roe-type Riemann solver is employed to calculate the common fluxes on open boundary faces

$$\boldsymbol{F}_n^{com} = \frac{1}{2}\left[\boldsymbol{F}_n^L + \boldsymbol{F}_n^R - (\boldsymbol{Q}^R - \boldsymbol{Q}^L)\widehat{\mathbb{R}}|\widehat{\boldsymbol{\Lambda}}|\widehat{\mathbb{L}}\right], \tag{26}$$

where flux $\mathbf{F}_n$ and unknown vector $\mathbf{Q}$ on left and right sides are obtained from their interior solution values and user-specified ambient conditions, respectively. Symbol ^denotes the arithmetic averaging operator and symbol | | denotes the absolute value operator.

### IV.C. CBC formulation

The analysis of the eigensystem of the 1D GLM-MHD system is conducted first since only the 1D governing equations in the normal direction are considered along boundary faces. Figure 3 shows the relation between the global Cartesian coordinates system (x, y, z) and the local coordinates system (n, t1, t2). The transformation from local coordinates to its global counterpart is shown below,

$$\begin{bmatrix} x \\ y \\ z \end{bmatrix} = \overline{\mathbb{R}} \begin{bmatrix} n \\ t1 \\ t2 \end{bmatrix}, \tag{27}$$

where $\overline{\mathbb{R}}$ is the standard rotation matrix.

The 1D GLM-MHD system of governing equations in the global coordinates reference at open boundary faces can be expressed in the form below

$$\frac{\partial \boldsymbol{Q}}{\partial t} + \frac{\partial \boldsymbol{F}_n}{\partial n} = 0, \tag{28}$$

$$\frac{\partial \boldsymbol{W}}{\partial t} + \mathbb{A}_p\frac{\partial \boldsymbol{W}}{\partial n} = 0, \tag{29}$$

where $\mathbf{Q}$ is the vector of conservative unknown variables, $\mathbf{Q} = [\rho, \rho v_x, \rho v_y, \rho v_z, Bx, By, Bz, \rho E, \psi]^T$ , W is the vector of primitive unknown variables, $\mathbf{W} = [\rho, v_x, v_y, v_z, Bx, By, Bz, p, \psi]^T$ .

The 1D GLM-MHD system of governing equations in a local coordinates reference at open boundary faces can be expressed as below

$$\frac{\partial \boldsymbol{Q}'}{\partial t} + \frac{\partial \boldsymbol{F}_n'}{\partial n} = 0, \tag{30}$$

$$\frac{\partial \boldsymbol{W}'}{\partial t} + \mathbb{A}_p' \frac{\partial \boldsymbol{W}'}{\partial n} = 0, \tag{31}$$

where **Q**’ is the vector of conservative unknown variables in local coordinates, **Q**’ = $[\rho, \rho v_n, \rho v_{t1}, \rho v_{t2}, B_n, B_{t1}, B_{t2}, \rho E, \psi]^T$, **W**’ is the vector of primitive unknown variables in local coordinates, **W**’ = $[\rho, v_n, v_{t1}, v_{t2}, B_n, B_{t1}, B_{t2}, p, \psi]^T$.

After multiplying $\frac{\partial \boldsymbol{Q}'}{\partial \boldsymbol{W}'}$ on both sides of Equation (31), one can obtain

$$\frac{\partial \boldsymbol{Q}'}{\partial t} + \frac{\partial \boldsymbol{Q}'}{\partial \boldsymbol{W}'} \mathbb{A}_p' \frac{\partial \boldsymbol{W}'}{\partial \boldsymbol{Q}'} \frac{\partial \boldsymbol{Q}'}{\partial n} = 0. \tag{32}$$

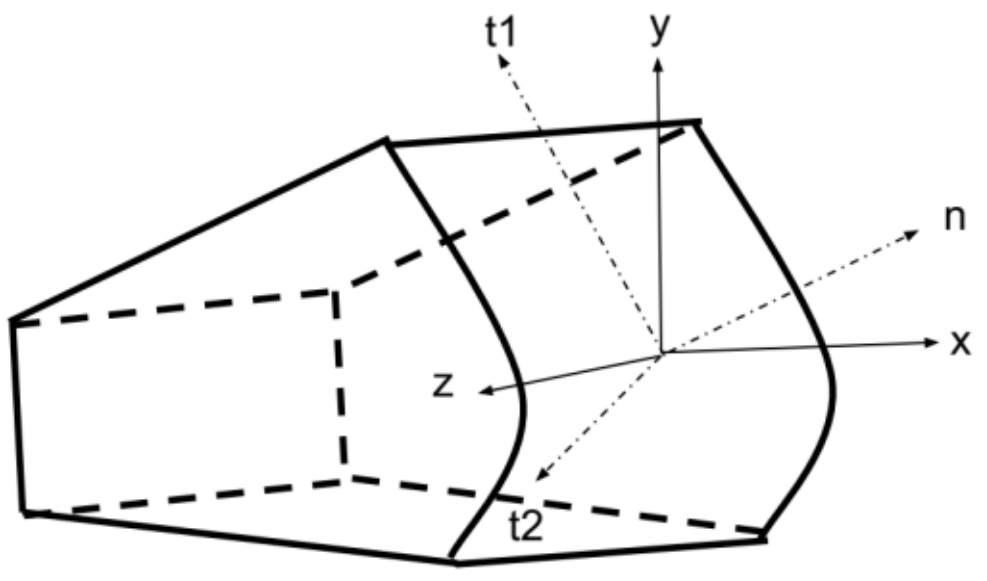

**Figure 3. Transformation between global Cartesian coordinates and local coordinates with respect to the curved boundary face.**

$\mathbb{A}_p'$ can be diagonalized as

$$\mathbb{A}_p' = \mathbb{R}_p' \boldsymbol{\Lambda} \mathbb{L}_p'. \tag{33}$$

The left and right eigenvector matrices expressed in terms of the conservative variables can be defined as

$$\begin{aligned} \mathbb{R}' &= \frac{\partial \boldsymbol{Q}'}{\partial \boldsymbol{W}'} \mathbb{R}_p', \\ \mathbb{L}' &= \mathbb{L}_p' \frac{\partial \boldsymbol{W}'}{\partial \boldsymbol{Q}'}. \end{aligned} \tag{34}$$

One can also define an extended rotation matrix as

$$\bar{\bar{\mathbb{R}}} = \begin{bmatrix} 1 & & & & \\ & \bar{\mathbb{R}} & & & \\ & & \bar{\mathbb{R}} & & \\ & & & 1 & \\ & & & & 1 \end{bmatrix}. \tag{35}$$

After substituting Equations (33) and (34) into Equation (32) and multiplying $\bar{\bar{\mathbb{R}}}$ on both sides of Equation (32), one can obtain

$$\frac{\partial \boldsymbol{Q}}{\partial t} + \overline{\overline{\mathbb{R}}}\mathbb{R}'\boldsymbol{\Lambda}\mathbb{L}'\overline{\overline{\mathbb{R}}}^{-1}\frac{\partial \boldsymbol{Q}}{\partial n} = 0. \tag{36}$$

Comparing Equation (36) to Equation (30), one can establish a new equivalence

$$\frac{\partial \boldsymbol{F}_n}{\partial n} = \mathbb{R}\boldsymbol{\Lambda}\mathbb{L}\frac{\partial \boldsymbol{Q}}{\partial n}, \tag{37}$$

where $\mathbb{R} = \overline{\overline{\mathbb{R}}}\mathbb{R}'$ and $\mathbb{L} = \mathbb{L}'\overline{\overline{\mathbb{R}}}^{-1}$. The 1D GLM-MHD system has nine eigenvalues and thus the right and left eigenvector matrices expressed in primitive variables have the following forms

$$\begin{aligned}\mathbb{R}' &= [\boldsymbol{r}_1', \boldsymbol{r}_2', \boldsymbol{r}_3', \boldsymbol{r}_4', \boldsymbol{r}_5', \boldsymbol{r}_6', \boldsymbol{r}_7', \boldsymbol{r}_8', \boldsymbol{r}_9'],\\ \mathbb{L}' &= [(\boldsymbol{l}_1')^T, (\boldsymbol{l}_2')^T, (\boldsymbol{l}_3')^T, (\boldsymbol{l}_4')^T, (\boldsymbol{l}_5')^T, (\boldsymbol{l}_6')^T, (\boldsymbol{l}_7')^T, (\boldsymbol{l}_8')^T, (\boldsymbol{l}_9')^T]^T.\end{aligned}$$

The characteristic vector is defined as

$$\mathscr{L} = \begin{bmatrix}\mathscr{L}_1\\ \mathscr{L}_2\\ \mathscr{L}_3\\ \mathscr{L}_4\\ \mathscr{L}_5\\ \mathscr{L}_6\\ \mathscr{L}_7\\ \mathscr{L}_8\\ \mathscr{L}_9\end{bmatrix} = \boldsymbol{\Lambda}\mathbb{L}\frac{\partial \boldsymbol{Q}}{\partial n}.$$

For open boundaries, the value of zero for an entry of $\mathscr{L}$ corresponds to an incoming wave. On the other hand, the value of $\lambda_i \boldsymbol{l}_i^T \frac{\partial \boldsymbol{Q}}{\partial n}$ for an entry of $\mathscr{L}$ corresponds to an outgoing wave where the formulations of $\lambda_i$ and $\boldsymbol{l}_i$ are detailed in the appendix.

### IV.D. Open Boundary Implementation in the FR Method

Without loss of generality, one can assume $\xi$ is the normal direction of a boundary face. The relation between $\frac{\partial \widetilde{\boldsymbol{F}}}{\partial \xi}$ and $\frac{\partial \boldsymbol{F}_n}{\partial n}$ on boundary face is given as follows,

$$\begin{aligned}\frac{\partial \widetilde{\boldsymbol{F}}}{\partial \xi} &= \frac{\partial(|\mathcal{J}|\xi_x \boldsymbol{F} + |\mathcal{J}|\xi_y \boldsymbol{G} + |\mathcal{J}|\xi_z \boldsymbol{H})}{\partial \xi}\\ &= |\mathcal{J}|\frac{\partial \boldsymbol{F}_n}{\partial n} + \frac{\partial|\mathcal{J}|}{\partial \xi}\xi_x \boldsymbol{F} + \frac{\partial|\mathcal{J}|}{\partial \xi}\xi_y \boldsymbol{G} + \frac{\partial|\mathcal{J}|}{\partial \xi}\xi_z \boldsymbol{H}.\end{aligned} \tag{38}$$

Therefore, for the CBC method, the normal derivative of flux can be expressed as $\frac{\partial \boldsymbol{F}_n}{\partial n} = \mathbb{R}\mathcal{L}$, while for ZND, one can simply set it to zero, i.e., $\frac{\partial \boldsymbol{F}_n}{\partial n} = 0$. After substituting the corresponding $\frac{\partial \boldsymbol{F}_n}{\partial n}$ into Equation (38), one can obtain the derivative of flux for the FR method in the computational domain as

$$\begin{aligned}\frac{\widetilde{\boldsymbol{F}}^{Rec}(\xi)}{\partial \xi} &= \frac{\widetilde{\boldsymbol{F}}(\xi)}{\partial \xi} + \left[\widetilde{\boldsymbol{F}}_L^{com} - \widetilde{\boldsymbol{F}}_i(0)\right]\frac{g_L(\xi)}{\partial \xi} + \left[\widetilde{\boldsymbol{F}}_R^{com} - \widetilde{\boldsymbol{F}}_i(1)\right]\frac{g_R(\xi)}{\partial \xi}\\ &= |\mathcal{J}|\frac{\partial \boldsymbol{F}_n}{\partial n} + \frac{\partial|\mathcal{J}|\xi_x}{\partial \xi}\boldsymbol{F}(\boldsymbol{Q}^L) + \frac{\partial|\mathcal{J}|\xi_y}{\partial \xi}\boldsymbol{G}(\boldsymbol{Q}^L) + \frac{\partial|\mathcal{J}|\xi_z}{\partial \xi}\boldsymbol{H}(\boldsymbol{Q}^L),\end{aligned} \tag{39}$$

where the only unknown variable is the common flux ($\widetilde{\boldsymbol{F}}_L^{com}$ or $\widetilde{\boldsymbol{F}}_R^{com}$) at the flux points along the boundary faces. Equation (39) can be solved. For the ARBC method, the common normal flux $\boldsymbol{F}_n^{com}$ is computed directly using Equation (26). The corresponding $\widetilde{\boldsymbol{F}}^{com}$ can be obtained after a trivial transformation, $\widetilde{\boldsymbol{F}}^{com} = |\nabla\xi|\boldsymbol{F}_n^{com}$.

# V. Numerical Tests

## V.A. 1D Perturbation Test

In this section, three different types of open boundary conditions are tested via a 1D problem with an initial perturbation imposed on the pressure and density fields, which generates a set of characteristic waves traveling towards the boundaries as shown in Figure 4.

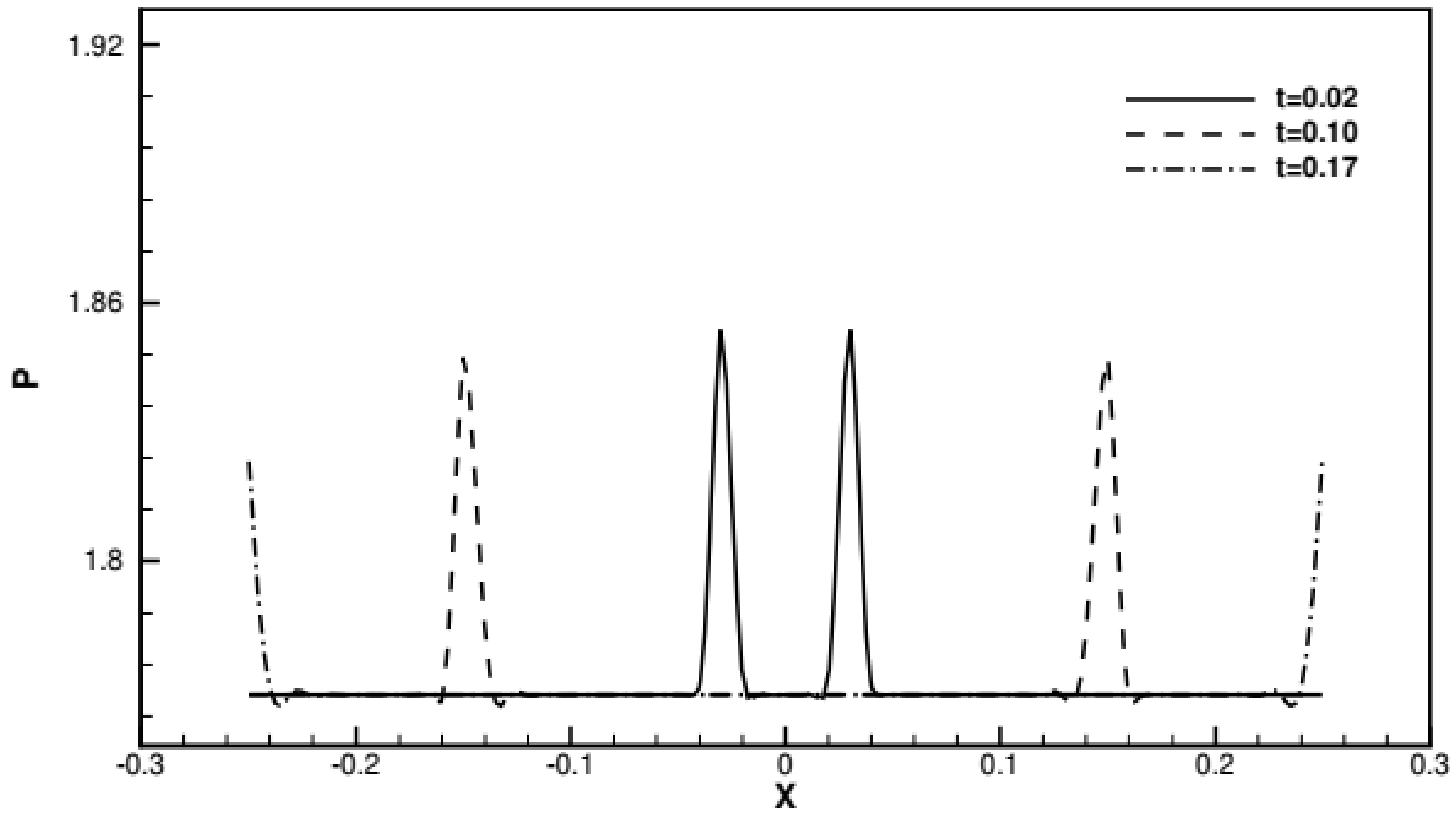

**Figure 4. Propagation of pressure wave at $A = 0.1$**

The initial conditions are specified as

$$\begin{cases} \rho(x,t=0) = \rho_\infty(1+Ae^{\frac{-x^2}{2\sigma^2}}) \\ \rho u(x,t=0) = \rho u_\infty \\ \rho v(x,t=0) = \rho v_\infty \\ \rho w(x,t=0) = \rho w_\infty \\ p(x,t=0) = p_\infty(1+Ae^{\frac{-x^2}{2\sigma^2}}) \\ Bx(x,t=0) = Bx_\infty \\ By(x,t=0) = By_\infty \\ Bz(x,t=0) = Bz_\infty \end{cases}, \qquad (40)$$

where $\rho_\infty = 1.368$, $u_\infty = w_\infty = 0.0$, $v_\infty = 1.0$, $p_\infty = 1.769$, $Bx_\infty = 1.0$, $By_\infty$=$Bz_\infty$=0.0, $\sigma$=0.004. Two different perturbation magnitudes of $A$=0.1 and $A$=1.0 are chosen representing small and large perturbations, respectively. The domain size ranges from $x$=-0.25 to $x$=0.25. A third-order accurate, strong-stability-preserving four-stage Runge-Kutta scheme introduced by Spiteri and Ruuth[8] is utilized for time marching.

Grid-independence evaluation is first performed using the third-order FR scheme. The results are shown in Figure 5. One can see that the density norm stays almost unchanged once the cell number reaches beyond 16.

Based on the grid independence analysis, the cell number is set to 32 ($\Delta$x = 0.0025) for the following study. Both the third and fourth order FR schemes are tested. The numerical solutions obtained from a larger domain ($-0.5 \leq x \leq 0.5$) is chosen as the reference solutions for error calculations and plotting. The L2 norms of density are calculated and shown in Figure 6. The results are starting from t = 0.14, at which the perturbation waves reach the two ends of the physical domain. From Figure 6(a), one can see that after the major perturbation wave leaves the whole domain at around t = 0.18, the L2 norm for the CBC method is smaller than the other two methods, which indicates that the CBC method allows the smallest amount of reflections of perturbation waves, while the ZND method generates the largest amount of reflections, since its L2 norm level is larger than of the other methods. The performance of the ARBC method in controlling reflections is the second best. Figure 6(a) also shows that this conclusion is independent of the order of accuracy since both the third and fourth order FR methods have been tested and show the same trend. Figure 6(b) shows the result for test cases with a large perturbation. It is shown that the L2 norm of density for each method is larger than the one with a small perturbation. This is reasonable because once the perturbation amplitude is 10 times larger, the refection amplitude should also increase. When comparing different methods with the same large perturbation, one can easily make a similar conclusion as the cases with small perturbations, i.e., the CBC method generates the lowest level of reflections, the ARBC method is the second best, whereas the ZND method produces the highest level of reflection.

The perturbation amplitude is further increased to A = 1.5. Table 1 shows that only the CBC method can get converged results while both ZND and ARBC methods diverged during simulations. Figure 7 shows the L2 norm of density for the CBC method at A = 1.5. However, for the ZND and ARBC methods, once the perturbation waves hit the open boundary faces, both MPI-Fortran runs blew up with the occurrence of NaN numbers. This result indicates that the CBC method is more robust than ZND and ARBC. The main reason lies in that the CBC method is the only one in three candidates that fully considers the directions of different characteristic waves, while the ZND method assumes all waves are going outward and the ARBC method uses some arithmetic averaging for calculating eigenvectors and eigenvalues as shown in Equation (26) to approximate all characteristic waves.

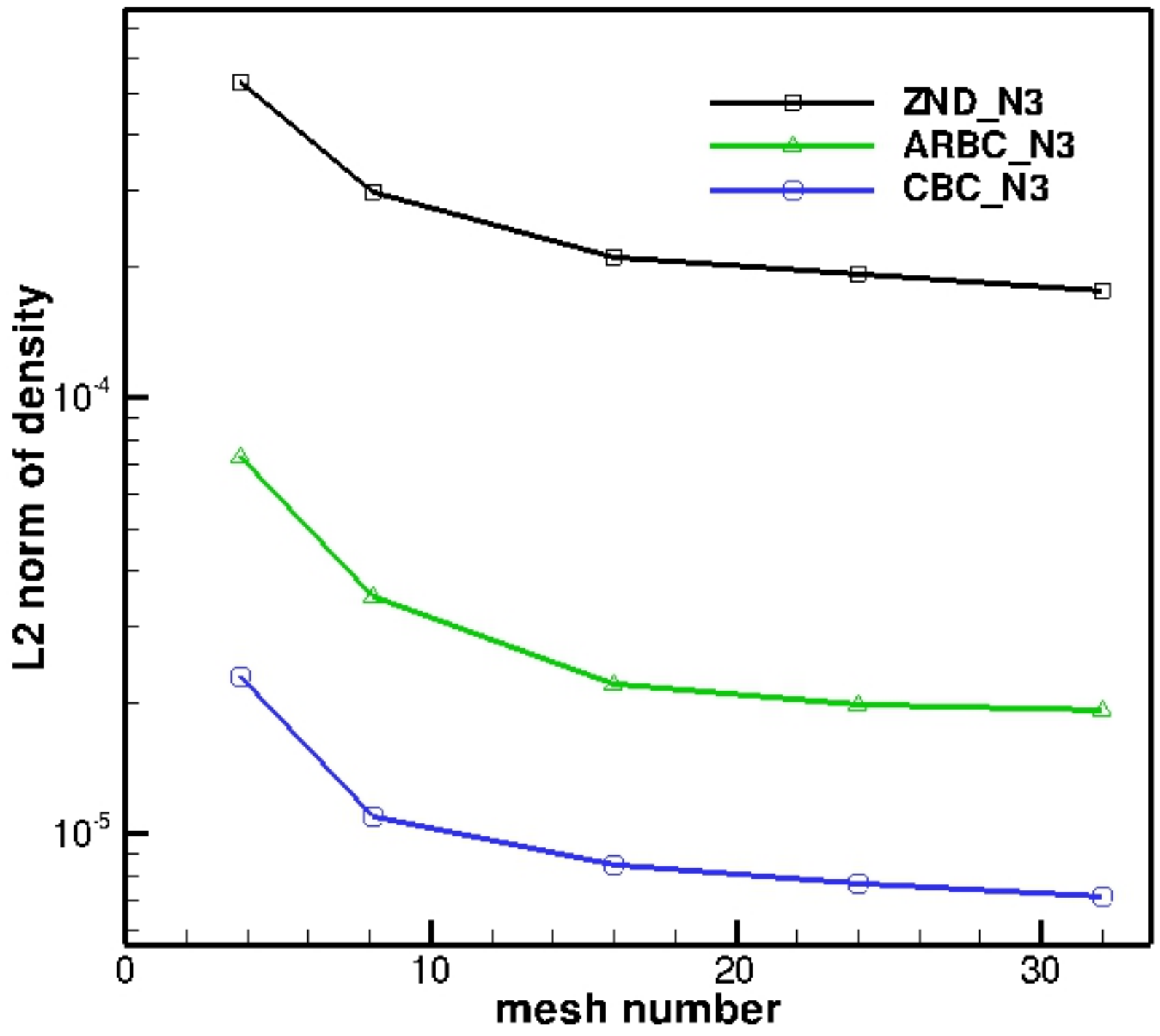

**Figure 5. Grid independence at $A = 0.1$**

### V.B. 2D Perturbation Test

A 2D perturbation test problem was investigated to evaluate the performance gain of our proposed CBC method over ZND and ARBC methods. The initial condition is the same as the 1D case shown in Equation (40) except that the x variable in the power of $e$ is replaced with radius r. The physical domain is a square box, $-0.25 \le x, y \le 0.25$. The grid spacing in both directions is set to Δx = Δy = 0.03125. The perturbation is initialized around the origin point. The prescribed perturbation waves will travel toward the boundaries as shown in Figure 8. The L2 norms of density are calculated and compared among different boundary conditions. Figure 9 shows results with the third-order FR scheme. Similarly, Figure 10 presents the L2 norms of density for the fourth-order FR scheme. Figure 8(a) shows the results with a small perturbation amplitude, A=0.1. One can see that the CBC method generates smaller reflection of perturbation waves bouncing back from the open boundary than ZND and ARBC methods. This result is consistent with the one obtained in the 1D test. However, when examining the performance of ZND and ARBC, one can see that after the perturbation waves are reflected back and forth several times, the L2 norm of density of the ARBC method exceeds that of the ZND method at around t = 0.6. This observation is slightly different from the results of the1D test case, where the L2 norms of density from the ARBC method are always lower than that of the ZND method, even though both methods exhibit L2 norms at the same order of magnitude. When looking at the results with a large perturbation amplitude, A=1.0, shown in Figure 8 (b), the ARBC method is slightly better than ZND, which is consistent with the results from the 1D test case. Most importantly, it is confirmed through this 2D case that the CBC method produces the smallest level of reflection for 2D simulations. It is also seen that the ZND and ARBC methods have comparable performances in controlling reflection with small perturbations, while for large perturbations, ARBC has better performance. For both 1D and 2D test cases, the CBC method performs better than ZND and ARBC, regardless of how large the perturbation is.

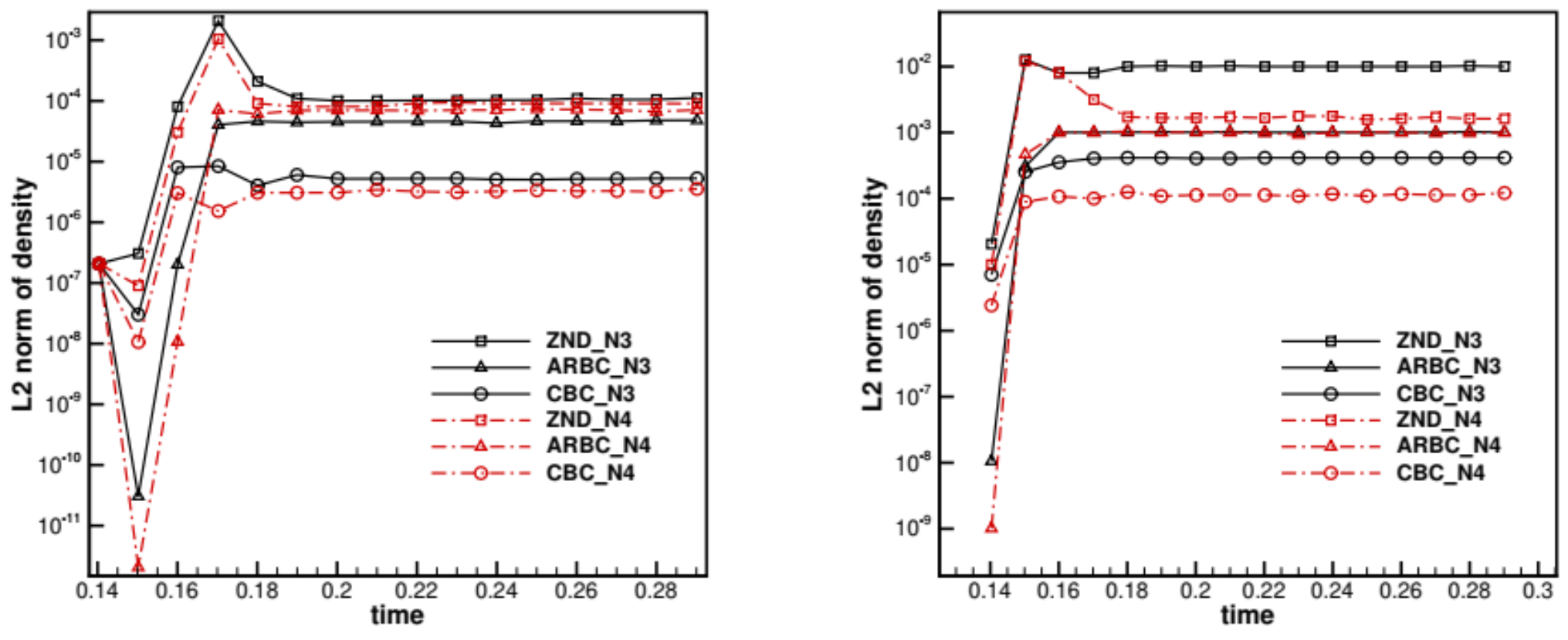

(a) A=0.1

(b) A=1.0

**Figure 6. L2 norm of density computed using different boundary conditions and two different orders of FR schemes**

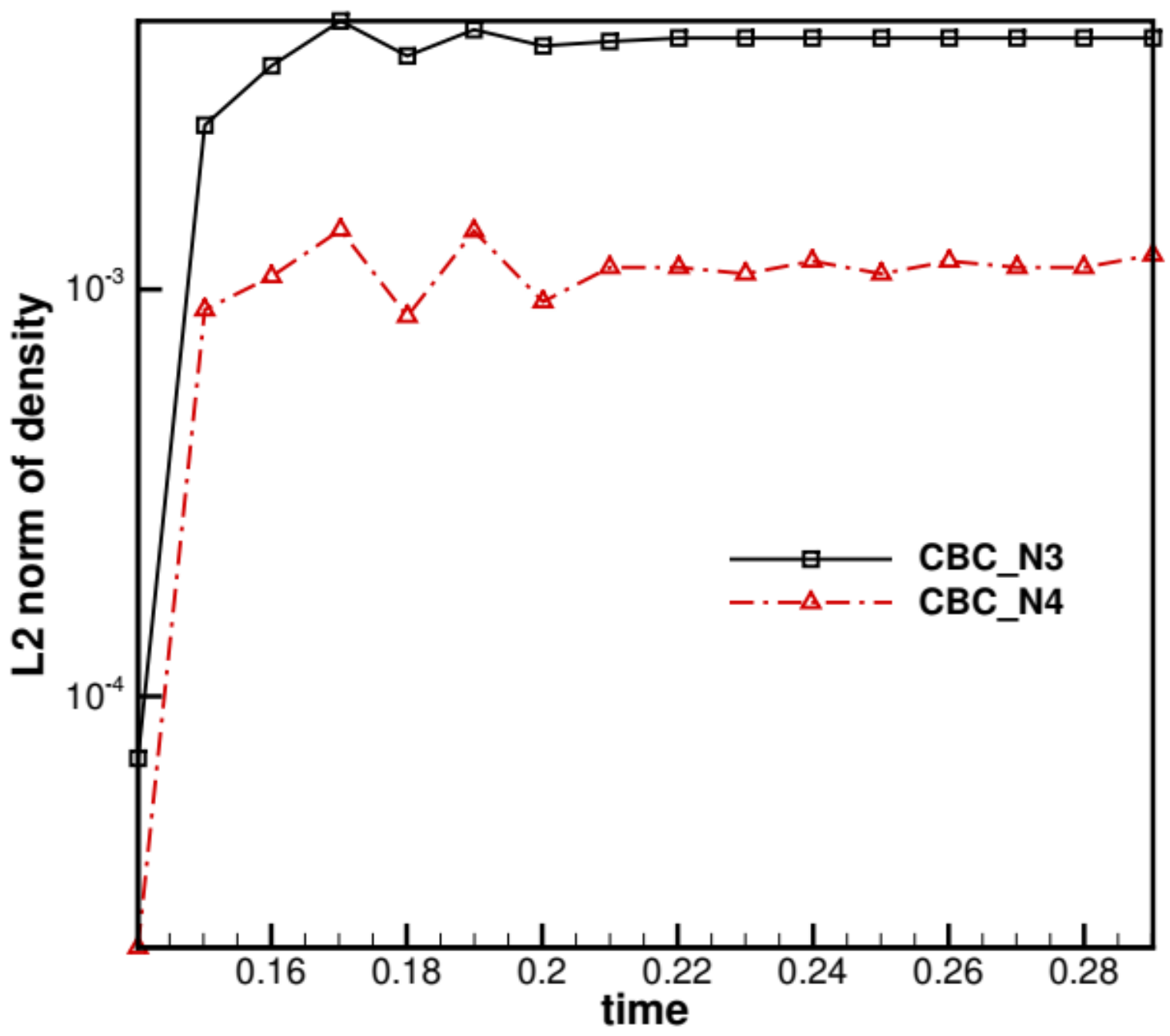

**Figure 7. L2 norm of density for CBC at** $A = 1.5$

| | A=0.1 | A=1.0 | A=1.5 |
|---|---|---|---|
| ARBC | √ | × | × |
| ZND | √ | √ | × |
| CBC | √ | √ | √ |

**Table 1. Robustness of the fourth-order FR scheme**

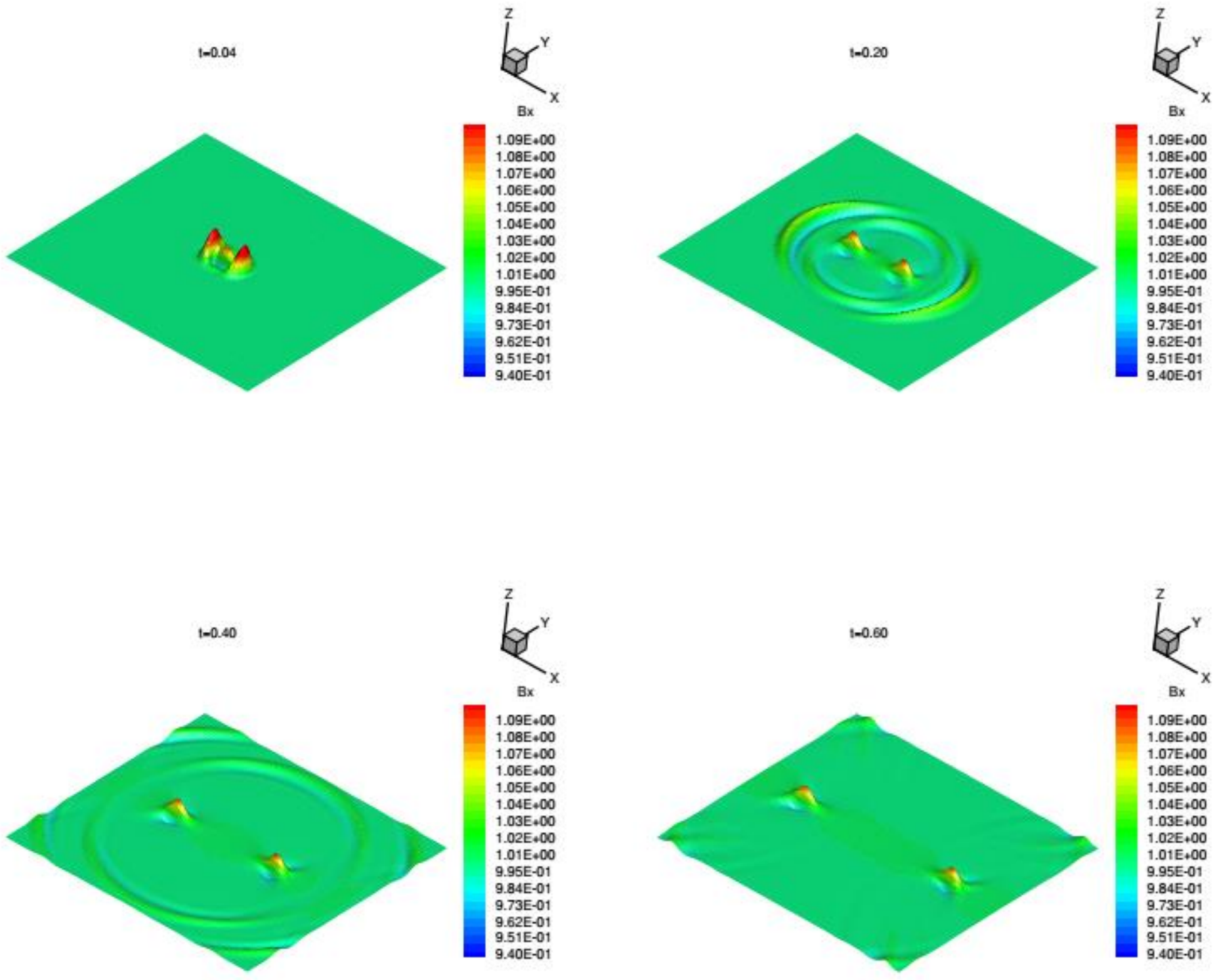

**Figure 8. Perturbed wave traveling at different time instants.**

## V.C. 3D Sphere and Spherical Shell Perturbation Tests

In this section two 3D perturbation test cases, a sphere and spherical shell, are conducted. Unstructured grids with all hexahedral cells are used. For the sphere test case, the radius at the outer boundary is 0.25. The initial perturbation is in the vicinity of the spherical center. The total number of grid cells used in this case is 7168. Figure 11 shows the mesh for the sphere test case where the perturbed wave travels outward from the spherical center. For the spherical shell test case, the radius at inner and outer boundaries are $r_{inner} = 0.25$ and $r_{outer} = 0.75$, respectively. The initial perturbation is in the middle range of the shell, i.e., $r = 0.5$. This specification allows the perturbation waves to travel both inward and outward as shown in Figure 12. The total number of grid cells is 30720. The third-order FR scheme is employed for both test cases.

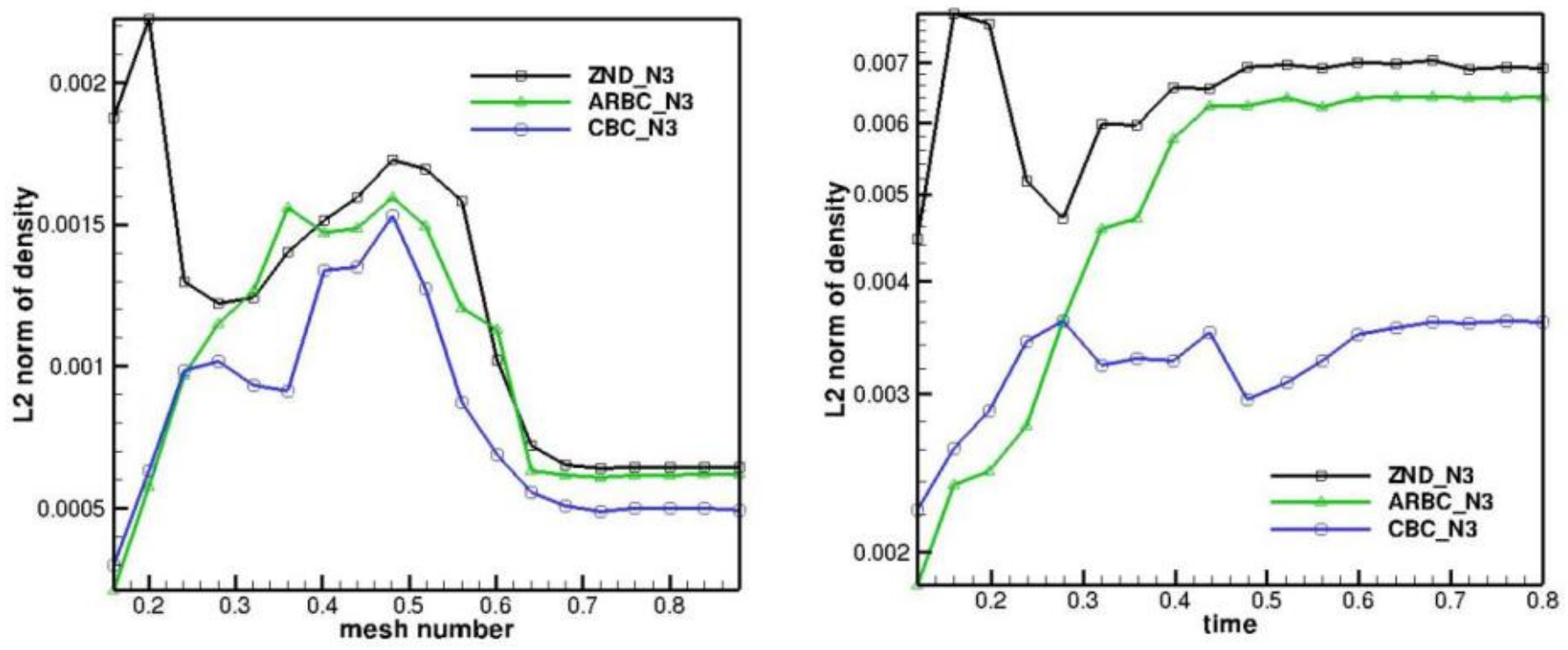

(a) A=0.1 (b) A=1.0

**Figure 9. L2 norms of density for 2D test problems computed using three different boundary conditions.**

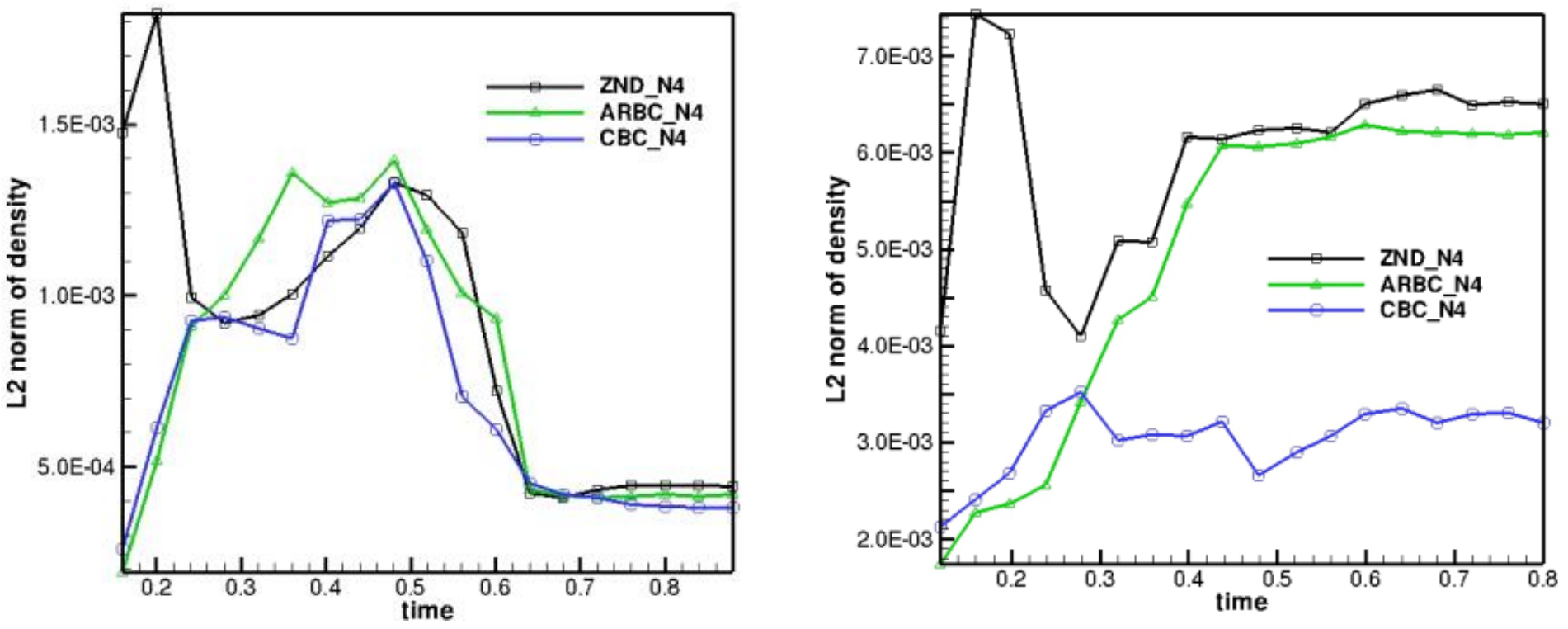

(a) A=0.1 (b) A=1.0

**Figure 10. L2 norms of density for 2D test problems computed using three different boundary conditions.**

Figures 13 and 14 show L2 norms of density for sphere and spherical shell test cases, respectively. For the sphere test case, one can see that for both small and large perturbation amplitudes, the L2 norms of density show similar trends, the CBC method generates the smallest level of reflection, the ZND method shows the highest level of reflection, whereas the ARBC is the middle. For the spherical shell test case with small perturbations, ZND and ARBC reach the same level of reflection after the major perturbation waves exit the computational domain ( $t\sim0.18$ ). In contrast, once increasing the perturbation amplitude to A=1.0, the difference gap between the reflection levels for ZND and ARBC keeps enlarging over the entire simulation period. Regardless of the different behaviors between ZND and ARBC, the CBC method produces the smallest level of reflection for both small and large perturbation cases.

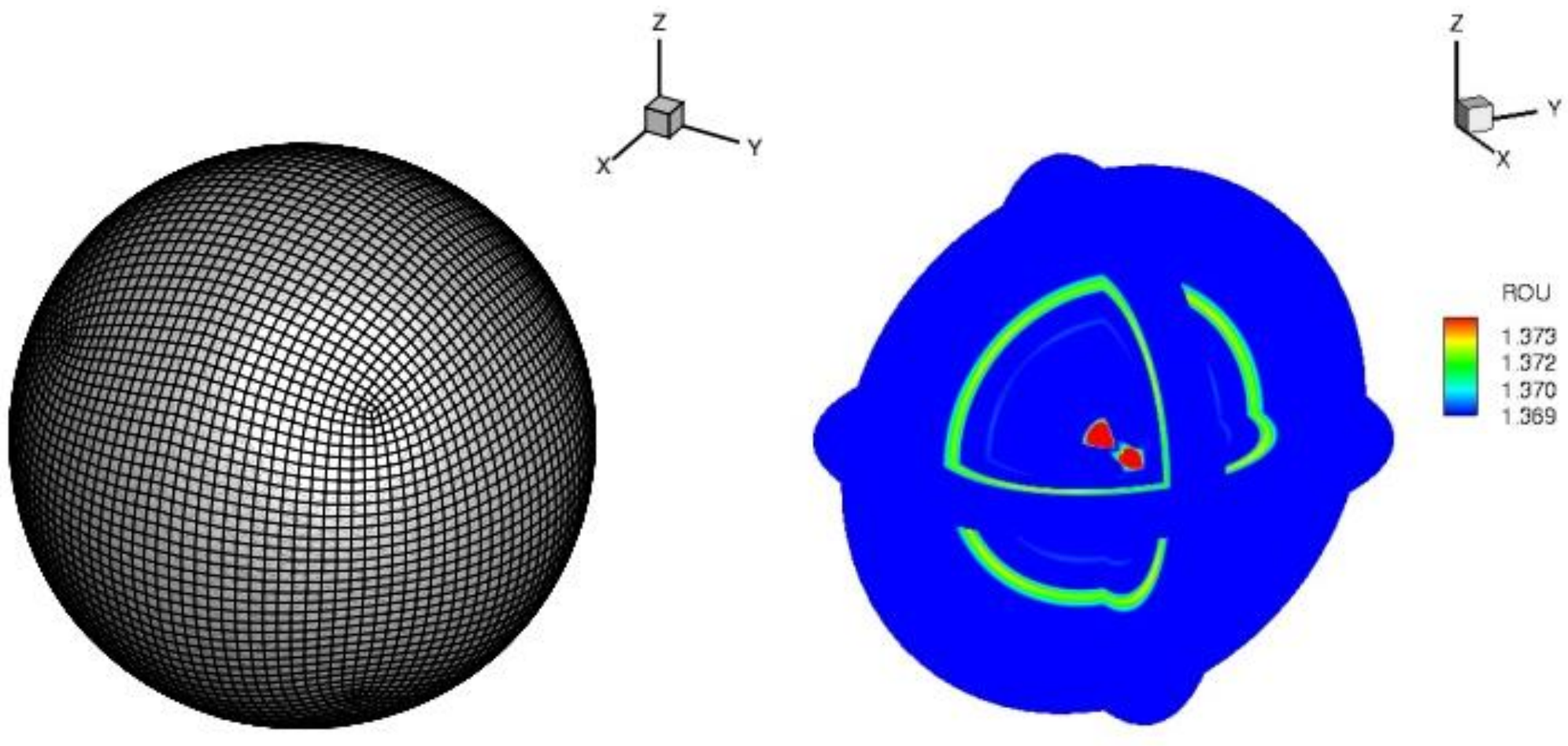

(a) Mesh for sphere test problem (b) Wave traveling

**Figure 11. Mesh for a test problem with sphere geometry and a snapshot for the outward traveling wave.**

## V.D. Magnetic Reconnection with Tearing Mode Instability

Magnetic reconnections occur in highly conducting plasmas where the magnetic topology breaks down and rearranges resulting in the conversion of magnetic energy to kinetic energy, thermal energy, and particle acceleration. It is normally an explosive event after a slow and gradual build-up process. Magnetic reconnection is thought to play a crucial role in eruptive solar events such as solar flares, coronal mass ejections, etc.

A major problem with magnetic reconnections in plasma physics is that the observed process of magnetic reconnection in the solar atmosphere is much faster than that predicted by theoretical models. For instance, solar flares proceed at several orders of magnitude faster than current theoretical predictions.

Over the past decades, researchers have proposed various theoretical models and conducted numerical simulations to explain the mechanics behind fast magnetic reconnections, including the classical Sweet-Parker model[9] and recent tearing-instability model[10–12].

Open boundaries are very useful in the simulations of magnetic reconnections. In order to capture small structures during the dynamic magnetic reconnection process, fine grid resolutions are required. Open boundaries can allow computational scientists to use truncated domain to simulate the whole process while simultaneously allocating more grid cells in targeted regions of interests and keeping the computational cost relatively low. Therefore, designing an accurate and stable open boundary is critical for magnetic reconnection simulations. In this section, The CBC boundary condition is applied to simulate the magnetic reconnections at a wide range of Lundquist numbers. ZND and ARBC boundary conditions were also employed for the purpose of comparison. Unfortunately, the simulations using ZND or ARBC failed and diverged at the very beginning stage of reconnection due to the accumulated contamination reflected from the open boundary faces. Both ZND and ARBC methods failed to accurately describe all characteristic waves as discussed in 1D perturbation test. Therefore, only the simulation results with the CBC method are discussed hereafter.

Various 2D numerical simulations have verified that steady Sweet-Parker model reconnection can be realized at low Lundquist numbers. It can, however, have tearing mode instability and then form a plasmoid once the Lundquist number exceeds a critical value of order $S{\sim}10^4$.[13–16] In this section the steady Sweet-Parker model reconnection at low Lundquist numbers is first verified and the tearing mode instability is also verified once the Lundquist number is increased beyond the critical values.

The initial conditions employed in this simulation is the classical Harris current sheet with small perturbations by setting $\rho = \rho_\infty(1 + \frac{1}{\beta cosh(x/a)^2})$, $p = \frac{\rho}{\gamma M}$, $u = v = w = 0, Bx = \frac{2\phi_0\pi}{L_y} sin(\frac{2\pi y}{L_y}) \cdot cos(\frac{\pi x}{L_x}), By = B_\infty tanh(\frac{x}{a}), Bz = 0$, where $\rho_\infty = \beta = 0.2$, $\beta$is the ratio of plasma pressure (p) to the magnetic pressure $(\boldsymbol{B} \cdot \boldsymbol{B}/(2\mu))$, $\gamma M^2 = 2$, and $\gamma$ is the ratio of specific heats. Finally, $M$ is Mach number, $B_\infty = 1.0$ is initial magnitude of magnetic field, $a = 0.5$ is the half width of current sheet, $\phi_0 = 0.1$ is the perturbation amplitude, $L_x = L_y = 20$ is the domain length along $x$ and $y$ directions, respectively. The grid spacing in both directions was set to $\Delta x = \Delta y = 0.4$. The fourth-order FR scheme was adopted along with the aforementioned Runge-Kutta method using a time step size of $dt = 5.0 \times 10^{-4}$. The domain size along x direction was sufficiently enlarged to ensure the boundary conditions over $x$ direction has little influence on reconnection region. Open boundary conditions along $x$ direction is set while periodic and symmetric boundary conditions were used along $y$ and z directions, respectively. Three different sets of resistivity coefficients were investigated, i.e., $\eta_c = 1.0 \times 10^{-3}$, $5.0 \times 10^{-3}$, $1.0 \times 10^{-2}$, corresponding to Lundquist numbers $S = 2.24 \times 10^4, 4.48 \times 10^3, 2.24 \times 10^3$ respectively. The Lundquist number was defined as $S = \frac{0.5 L_x c_a}{\eta_c}$ ,where $c_a = \frac{B_\infty}{\sqrt{\rho_\infty}}$.

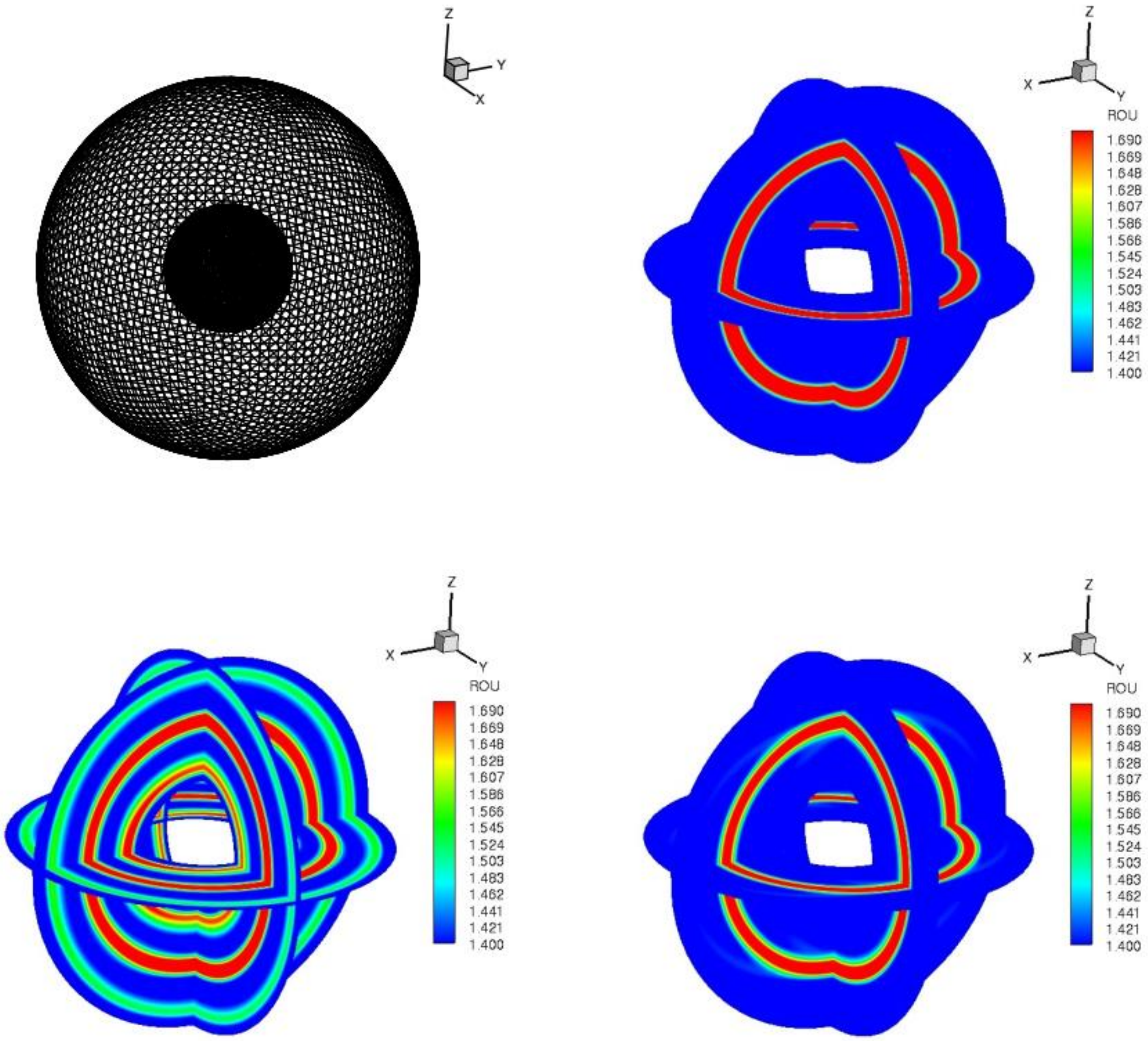

**Figure 12. Mesh for a test problem with spherical shell geometry and snapshots for inward and outward traveling waves.**

Figures 15, 16, and 17 show the evolution of current $Jz$ ($\boldsymbol{J} = \nabla \times \boldsymbol{B}/\mu$) for $\eta_c = 1 \times 10^{-2}, 5 \times 10^{-3}$, $1 \times 10^{-3}$, respectively. It is obvious that there is only steady Sweet-Parker reconnection happening at $\eta_c = 1 \times 10^{-2}$ and $5 \times 10^{-3}$, i.e., $S = 2.24 \times 10^3$ and $4.48 \times 10^3$. However, when decreasing the resistivity coefficient to $\eta_c = 1 \times 10^{-3}$, the tearing mode instability occurs. These results agree very well with the theoretical prediction of triggering point of tearing mode instability, which is $S{\sim}10^4$. Taking a further look at Figure 18, one can find that the nonlinear development stage at $\eta_c = 1 \times 10^{-3}$ starts at around $t = 35$. Before that it experiences a long linear development stage with notable differences in the slopes for different resistivity coefficients. The absolute values of curve slopes shown in Figure 18 are proportional to the reconnection rate $R$. The estimated reconnection rates $R$ are $\sim 0.154, \sim 0.285, \sim 0.375$ at $\eta_c = 1.0 \times 10^{-3}, 5.0 \times 10^{-3}, 1.0 \times 10^{-2}$, respectively. This result agrees well with the reconnection rate relation predicted by the Sweet-Parker model, $R{\sim}S^{-1/2}$.

In Figure 17, it is easy to notice that once the nonlinear phase starts, the m = 2 mode of plasmoids dominates the reconnection region as shown in the contour plot at t = 37.5. After that the two plasmoids merge with each other and form a m = 1 mode of plasmoid shown in the contour plot at t = 57.5. From t =50 to t=100, the reconnection process experiences a relatively stable period. After that, the primary plasmoid starts being ejected to one of the outflow regions and afterwards an elongated current sheet is formed again. During the plasmoid ejection process, the reconnection rate accelerates again following that relatively stable stage.

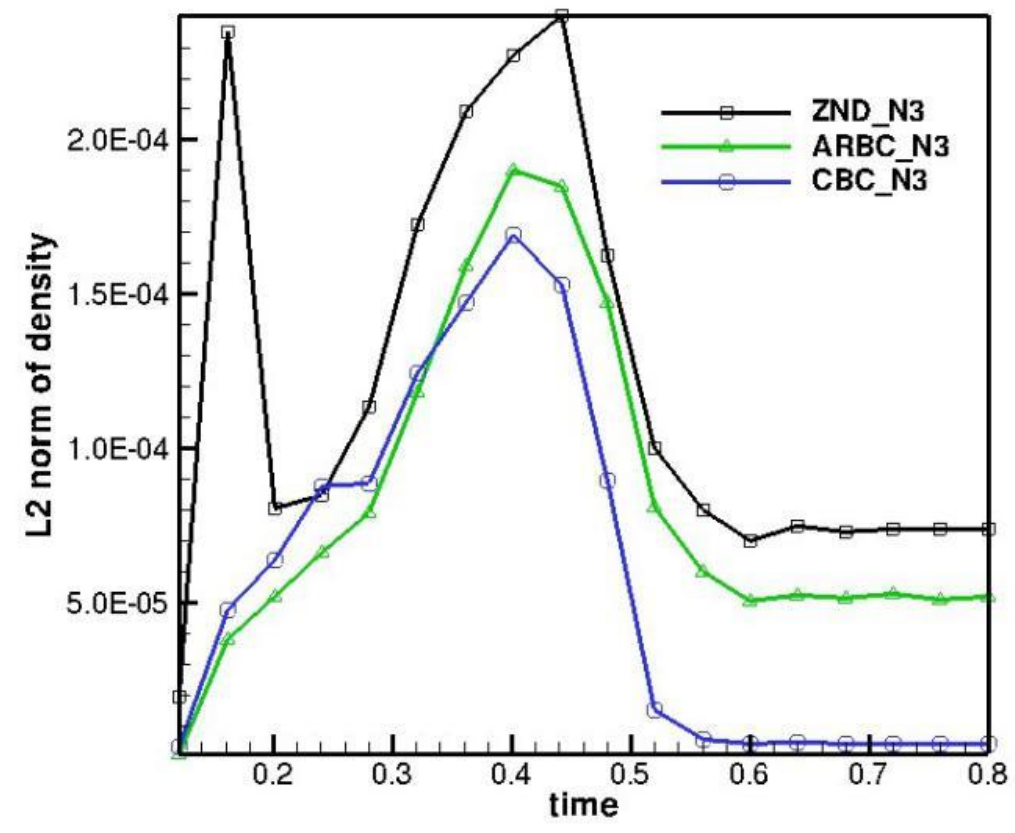

(a) A=0.1, N=3

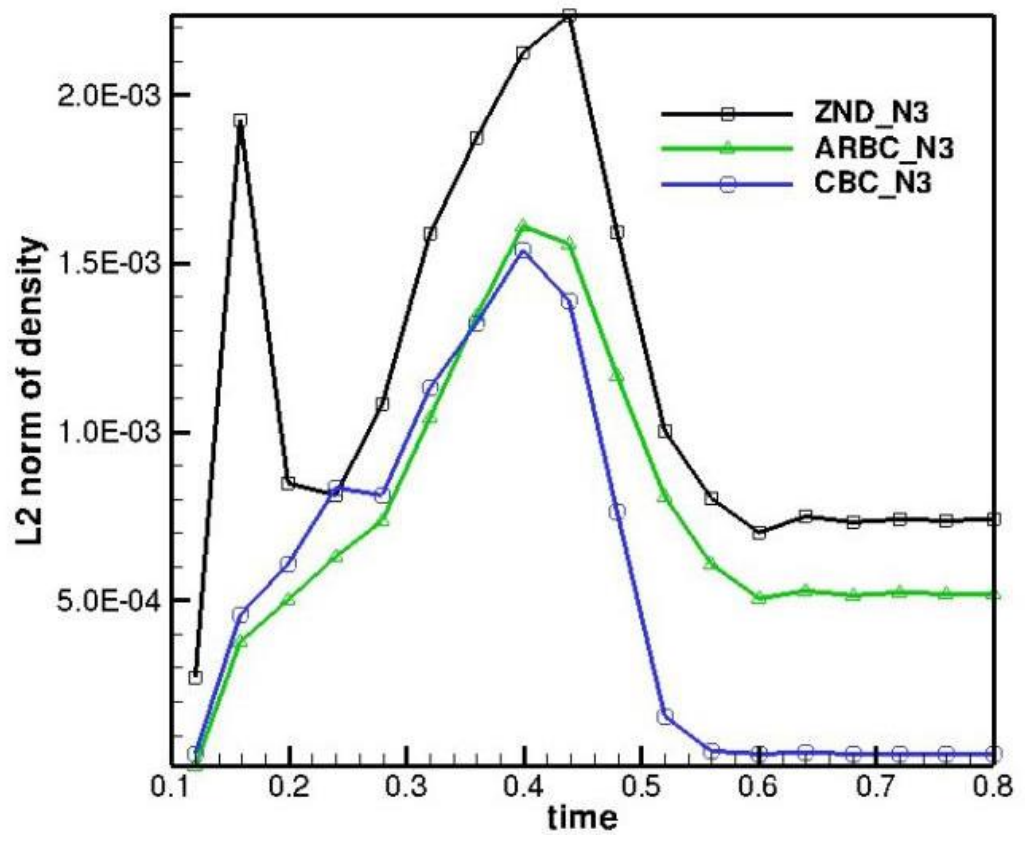

(b) A=1.0, N=3

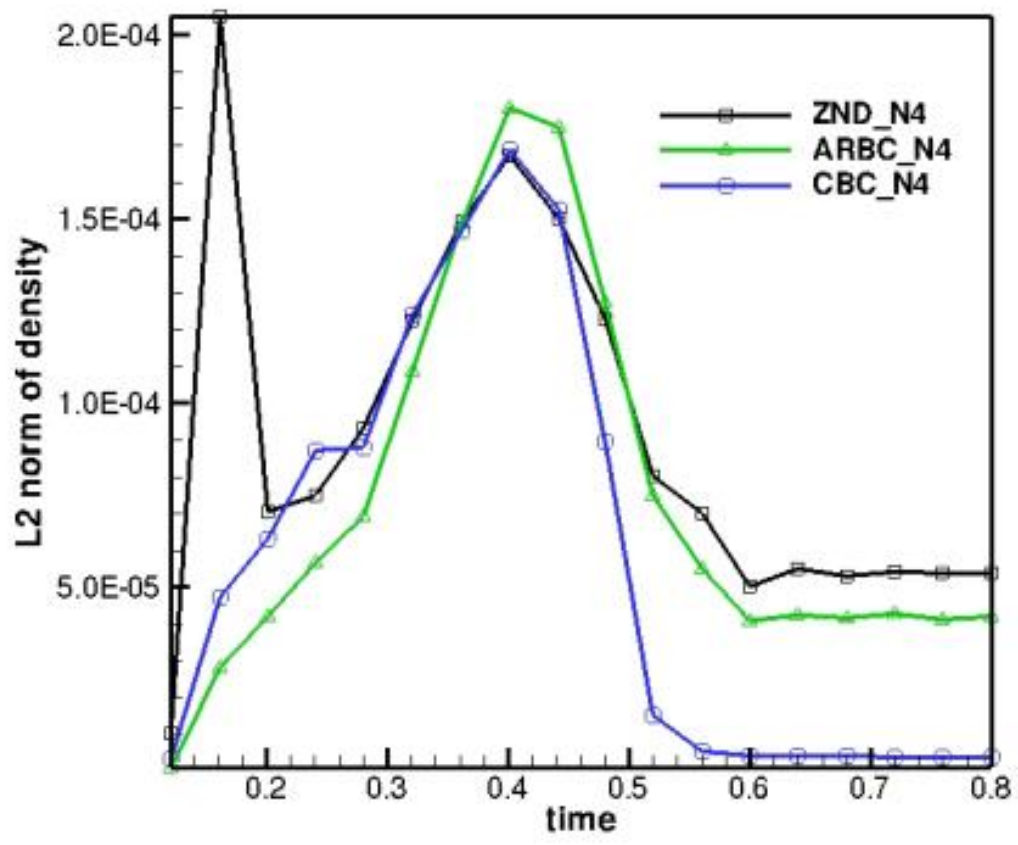

(c) A=0.1, N=4

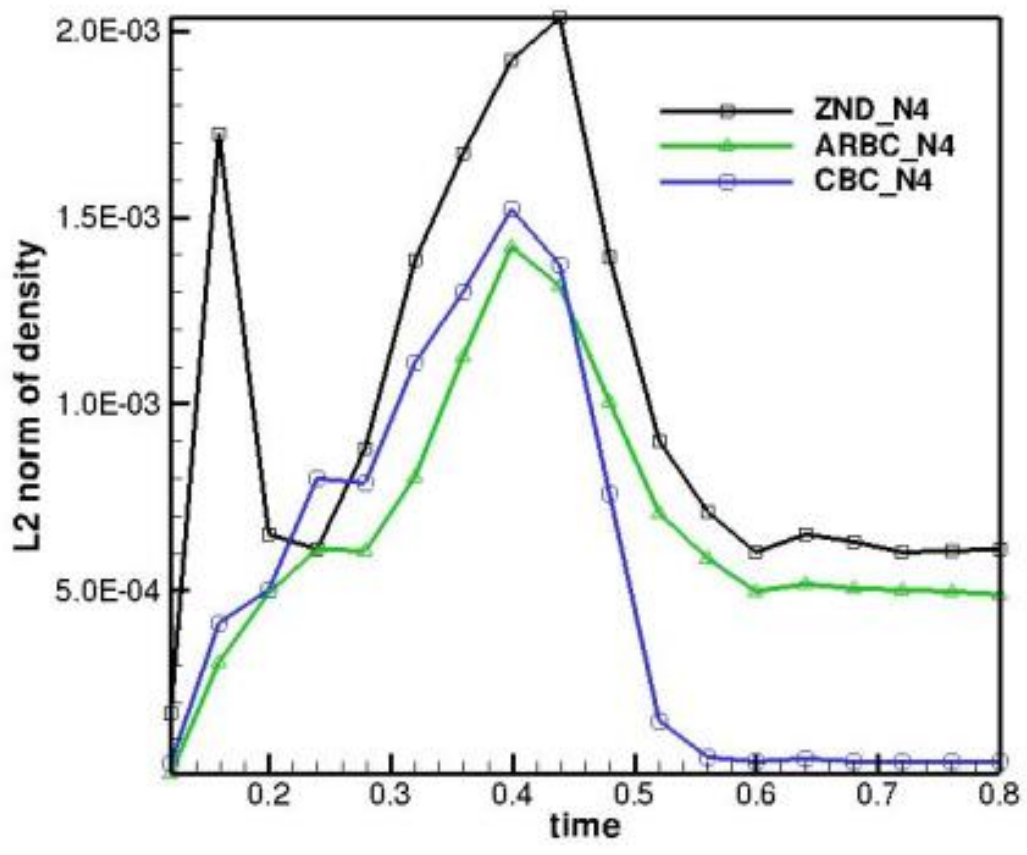

(d) A=1.0, N=4

**Figure 13. L2 norms of density for sphere test**

As shown in the above magnetic reconnection simulations, the Sweet-Parker current is indeed unstable once the Lundquist number exceeds a critical value. However, as discussed above, the theoretical scaling relation for the reconnection rate with S, i.e., $R\sim S^{-1/2}$ , has an intrinsic problem due to the contradictory fact that an infinitely large Lundquist number $S \rightarrow \infty$ would lead to infinitely fast instabilities. In an ideal MHD model, such a magnetic reconnection rate is impossible. The recent ideal tearing instability theory provides a new perspective that can resolve this paradoxical issue very nicely. For the purpose of demonstration, the ideal tearing mode instability is reproduced using the same test cases investigated by Del Zanna et al.[12] The results obtained from simulations in this paper agree very well with those provided in Del Zanna et al[12] at both linear and nonlinear stages. The initial conditions with a small perturbation on velocity fields are shown below,

$$\begin{cases} \rho = \rho_{\infty}(1+\frac{1}{\beta cosh(x/a)^2}), \quad u = \epsilon tanh(\xi)exp(-\xi^2)cos(ky) \\ v = \epsilon(2\xi tanh(\xi) - 1.0/cosh(\xi)^2)exp(-\xi^2)S^{1/2}sin(ky)/k, \quad w = 0 \\ p = \frac{1}{2}\rho, \quad Bx = 0, \quad By = B_{\infty}tanh(\frac{x}{a}), \quad Bz = 0 \end{cases} \tag{41}$$

where the pressure ratio is set to $\beta = 2.4$, the width of current sheet is $a = Ly/S^{1/3}$, the Lundquist number is set to $S = 1 \times 10^6, Ly = 1$, $\rho_\infty = \beta$, $B_\infty = 1.0$, the perturbation magnitude is specified as $\epsilon = 1.0 \times 10^{-3}$, $\xi = xS^{1/2}$, while the wave number is computed from $kLy = 2\pi m$, with $m = 10$. The periodic and open boundary conditions are chosen along x and y directions, respectively. The rectangular domain size is$[-20a, 20a] \times [0, Ly]$. The number of grid cells is $180 \times 450$. The fifth-order FR scheme is employed. The evolution of magnetic reconnection is shown in Figures 19 and 20. In the first snapshot, the process is still in the linear stage with $m = 3$ mode dominating the current sheet. When the tearing instability growth is over, the nonlinear phase sets in leading to further reconnection events and island coalescence as shown at the second and third snapshot. At this stage one can clearly observe the process leading to the creation of a single, large magnetic island as arising from coalescence. The whole process is very dynamic with the explosive creation of smaller and smaller islands. The small-scale islands then move towards the largest one, which is continually fed and agglomerating and thus continuously enlarges its size. Such a cascading explosive magnetic reconnection process is reminiscent of the flaring activity.

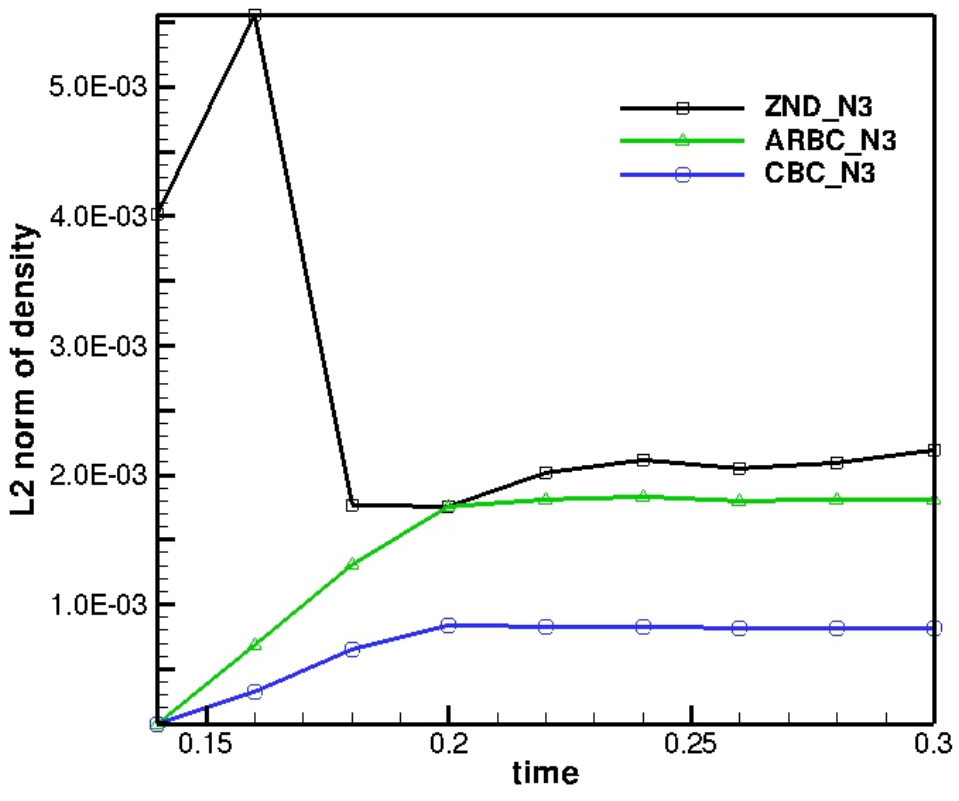

(a) A=0.1, N=3

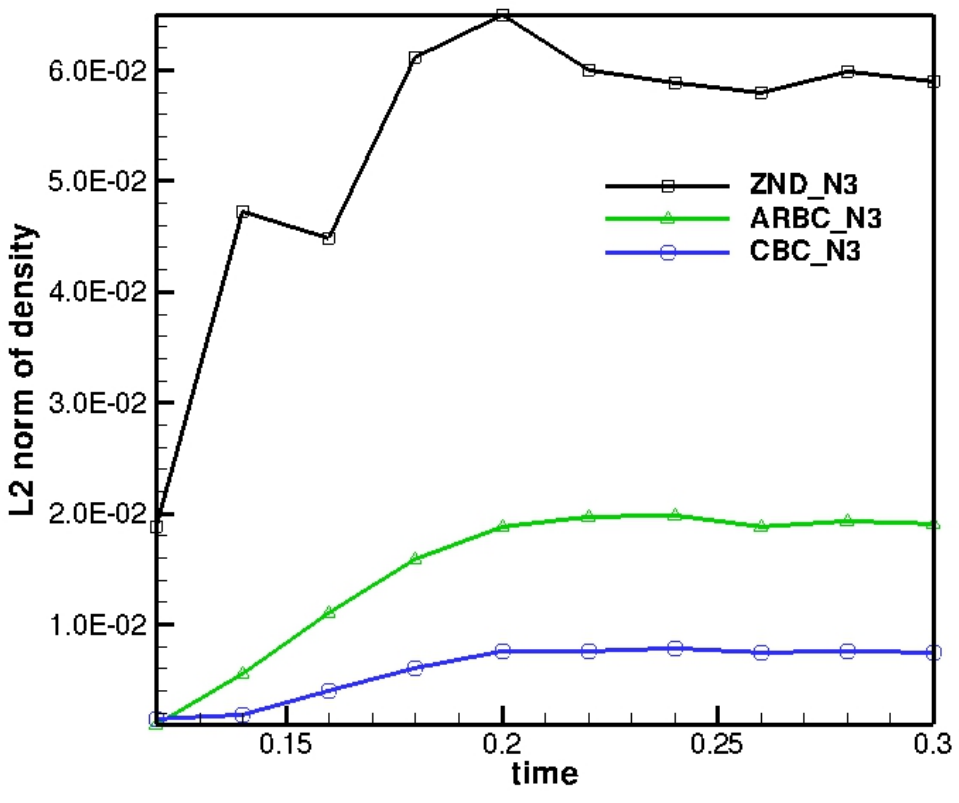

(b) A=1.0, N=3

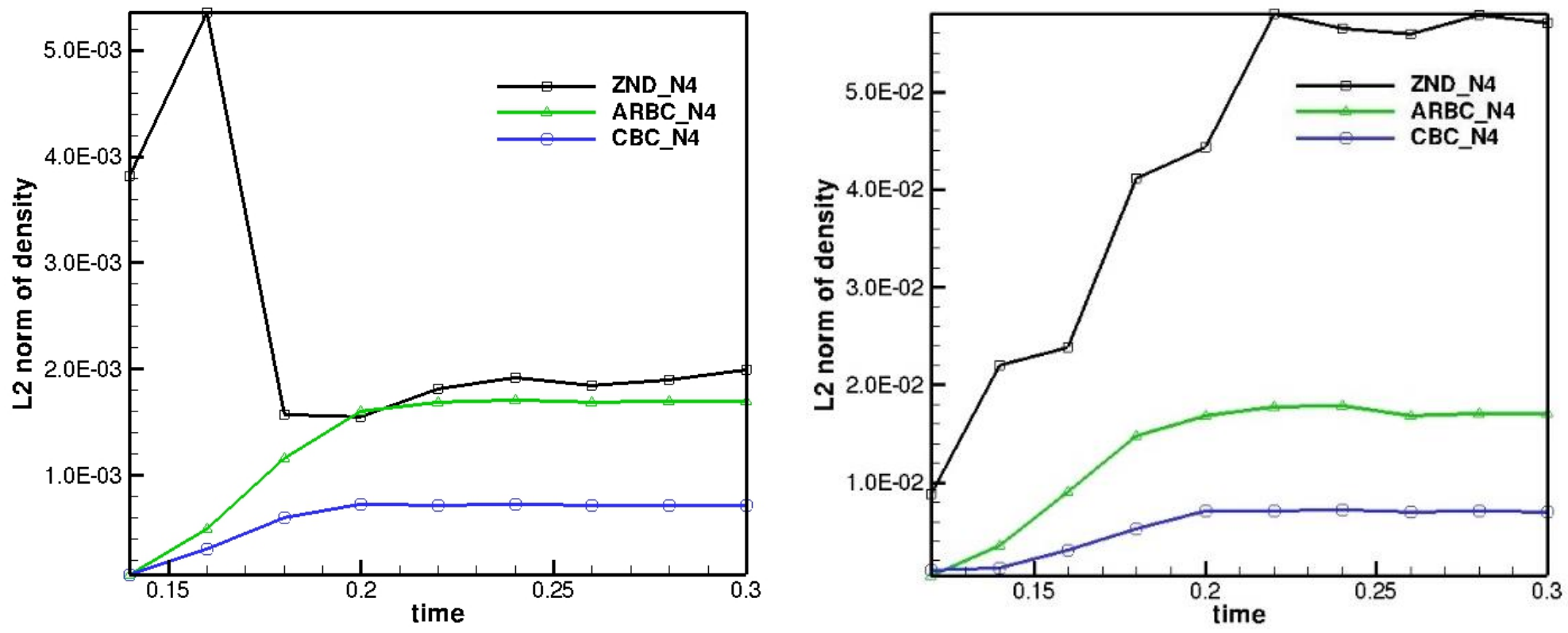

(c) A=0.1, N=4

(d) A=1.0, N=4

**Figure 14. L2 norms of density for the spherical shell test case**

# VI. Conclusions

For the first time, a characteristics-based boundary condition is implemented for solving the three-dimensional GLM-MHD system of equations on unstructured grids. For the purpose of comparison, two other boundary conditions, namely, zero normal derivative (ZND) and approximate Riemann solver boundary conditions (ARBC) are also implemented in the context of the Flux Reconstruction scheme. The performance of the FR method is evaluated and compared by computing 1D, 2D, and 3D perturbation tests using three different boundary conditions (CBC, ZND, and ARBC).

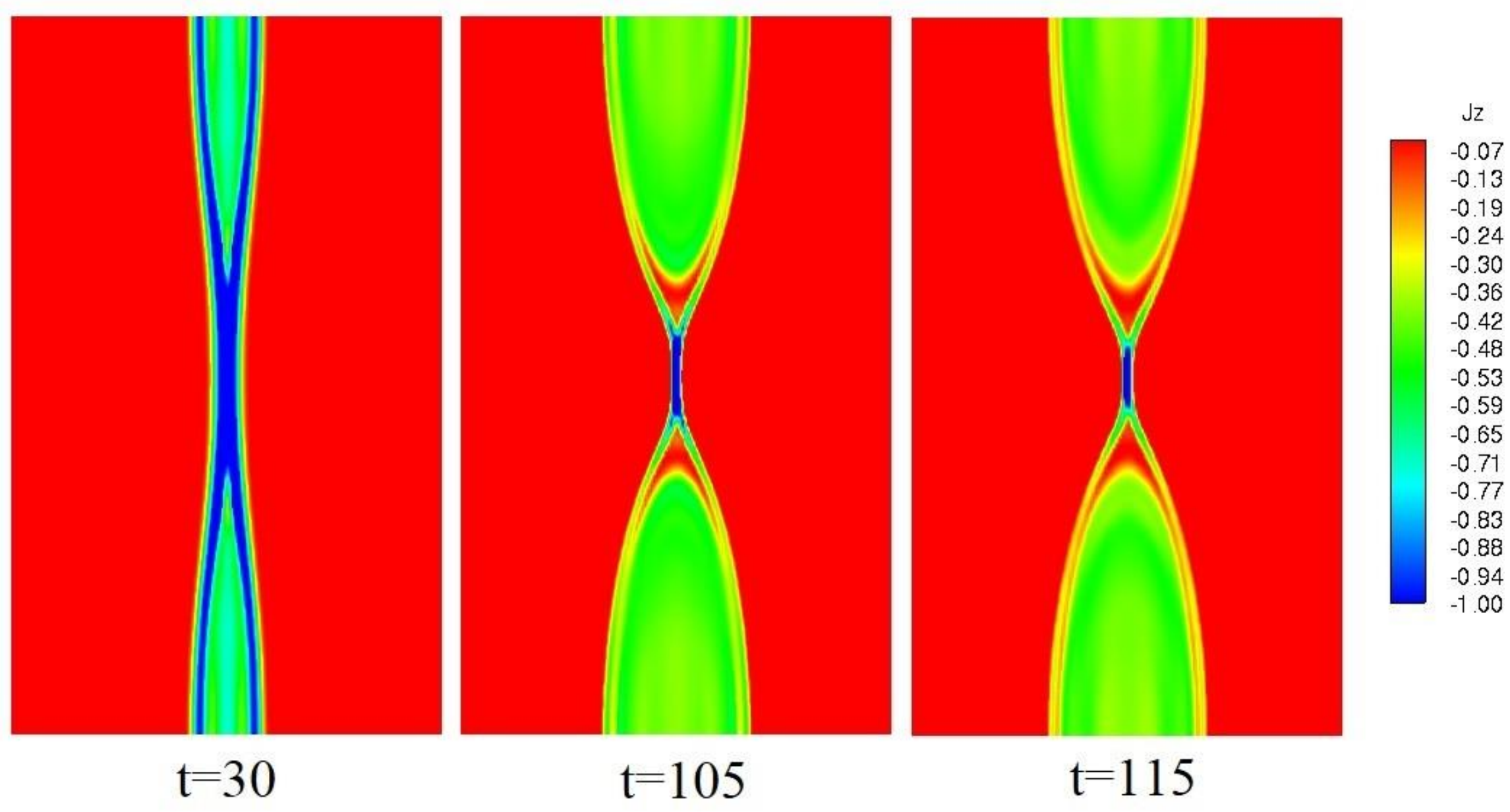

**Figure 15. Contour plot of current $Jz$ for $\eta_c = 1 \times 10^{-2}$, steady magnetic reconnection**

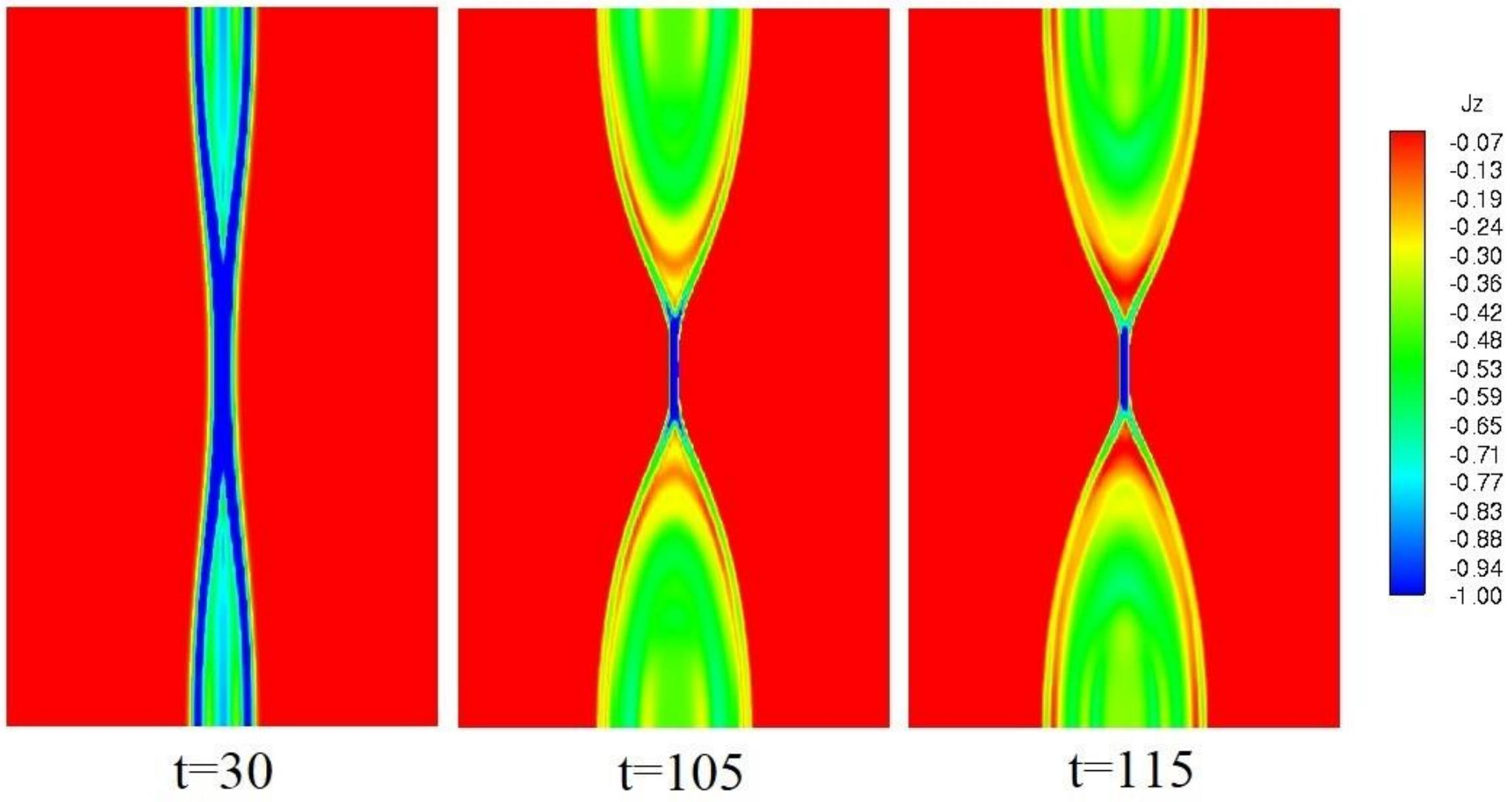

**Figure 16. Contour plots of current $Jz$ for steady magnetic reconnection problem with $\eta_c = 5 \times 10^{-3}$**

The ZND method has the simplest implementation, i.e., simply setting the normal flux derivative term on boundary faces to zero. However, numerical results demonstrate that ZND has notable reflections from the boundary faces, especially when waves of large amplitudes cross. The ARBC method involves certain approximate Riemann solvers to compute common flux values on the boundary faces. The directions of different characteristic waves are considered. Overall, ARBC outperforms ZND, especially when the amplitude of waves is moderate or large.

In the CBC method, the value of each characteristic variable is determined according to its propagating direction, which is more precise than traditional approximate Riemann solvers. Furthermore, the contribution of characteristic equations goes directly to the corrected term of the Flux Reconstruction scheme avoiding solving extra equations. Numerical results show that CBC generates smaller reflections from the boundaries than ZND and ARBC. More importantly, CBC is more robust when waves of large amplitudes propagate across the boundaries. The robustness of the CBC method is further demonstrated via the simulations of magnetic reconnections at increased Lundquist numbers. The tearing mode instability of magnetic reconnections is successfully realized by using the CBC method. However, both ZND and ARBC fail to predict the tearing mode due to instability issues.

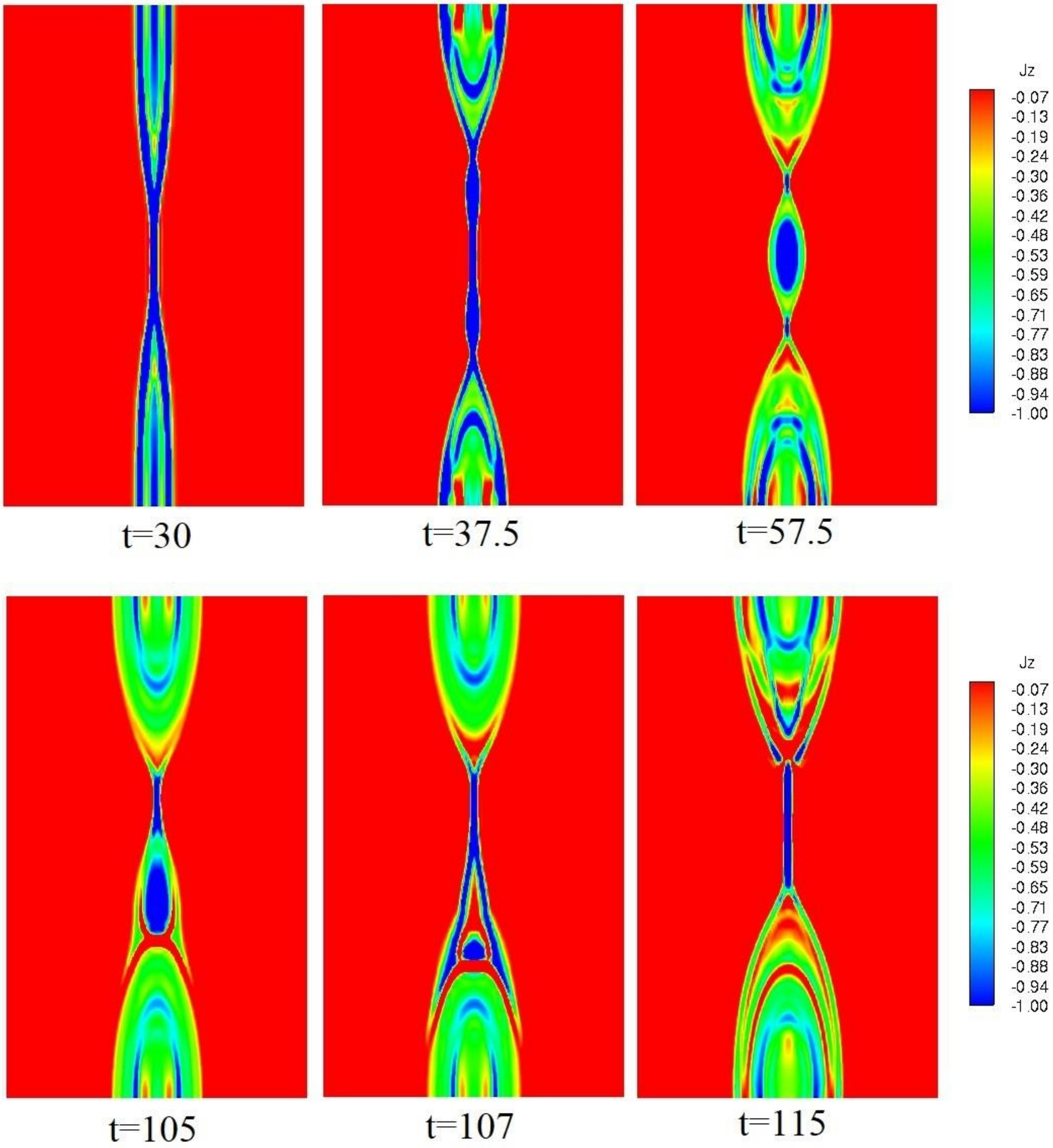

**Figure 17. Contour plots of current $Jz$ for unsteady magnetic reconnection problem with tearing mode and $\eta_c = 1 \times 10^{-3}$,**

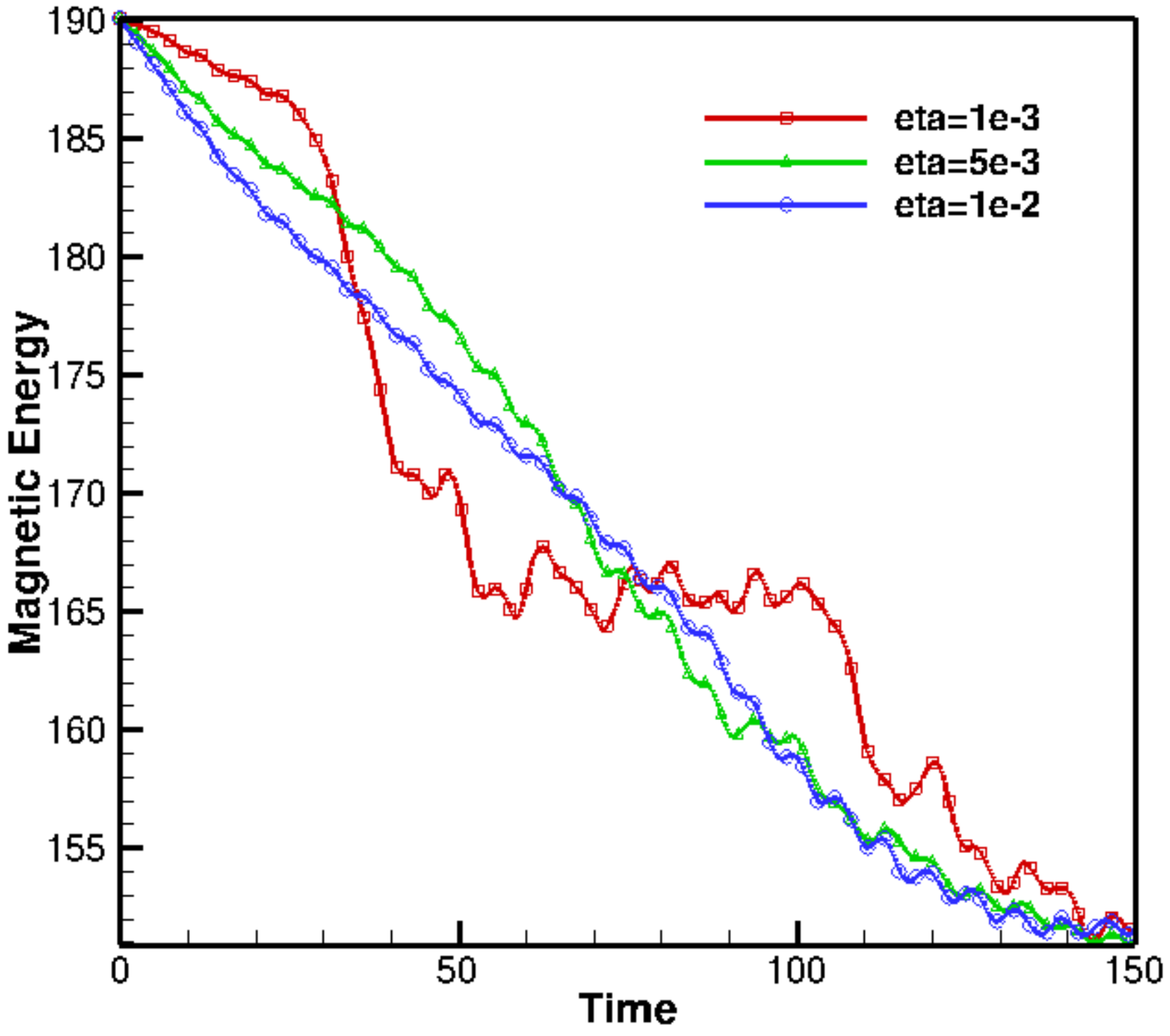

**Figure 18. Magnetic energy evolution over time**

# Acknowledgement

This research work was supported by a National Science Foundation (NSF) CAREER Award (No. 1952554) to Professor Chunlei Liang. Part of this CAREER grant was transferred from George Washington University to Clarkson University in 2019. The authors are grateful for the computing hours and technical support received from the NSF Extreme Science and Engineering Discovery Environment (XSEDE) team. Both authors would also like to acknowledge the financial support from the department of Mechanical and Aerospace Engineering and the School of Engineering and Applied Science at the George Washington University for Xiaoliang Zhang towards his Ph.D. degree.

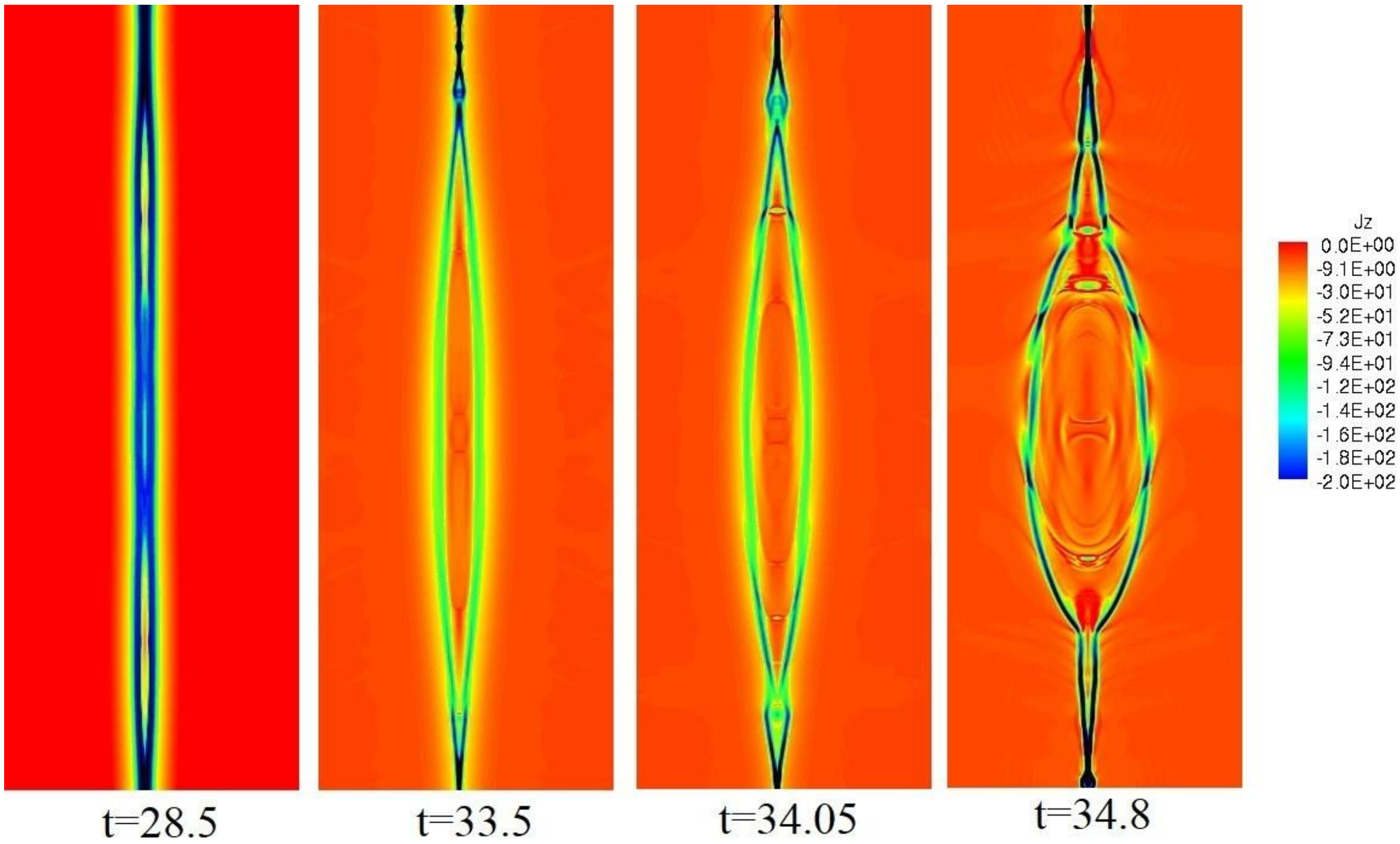

**Figure 19. Contour plots of $J_z$ for unsteady magnetic reconnection problem with ideal tearing mode**

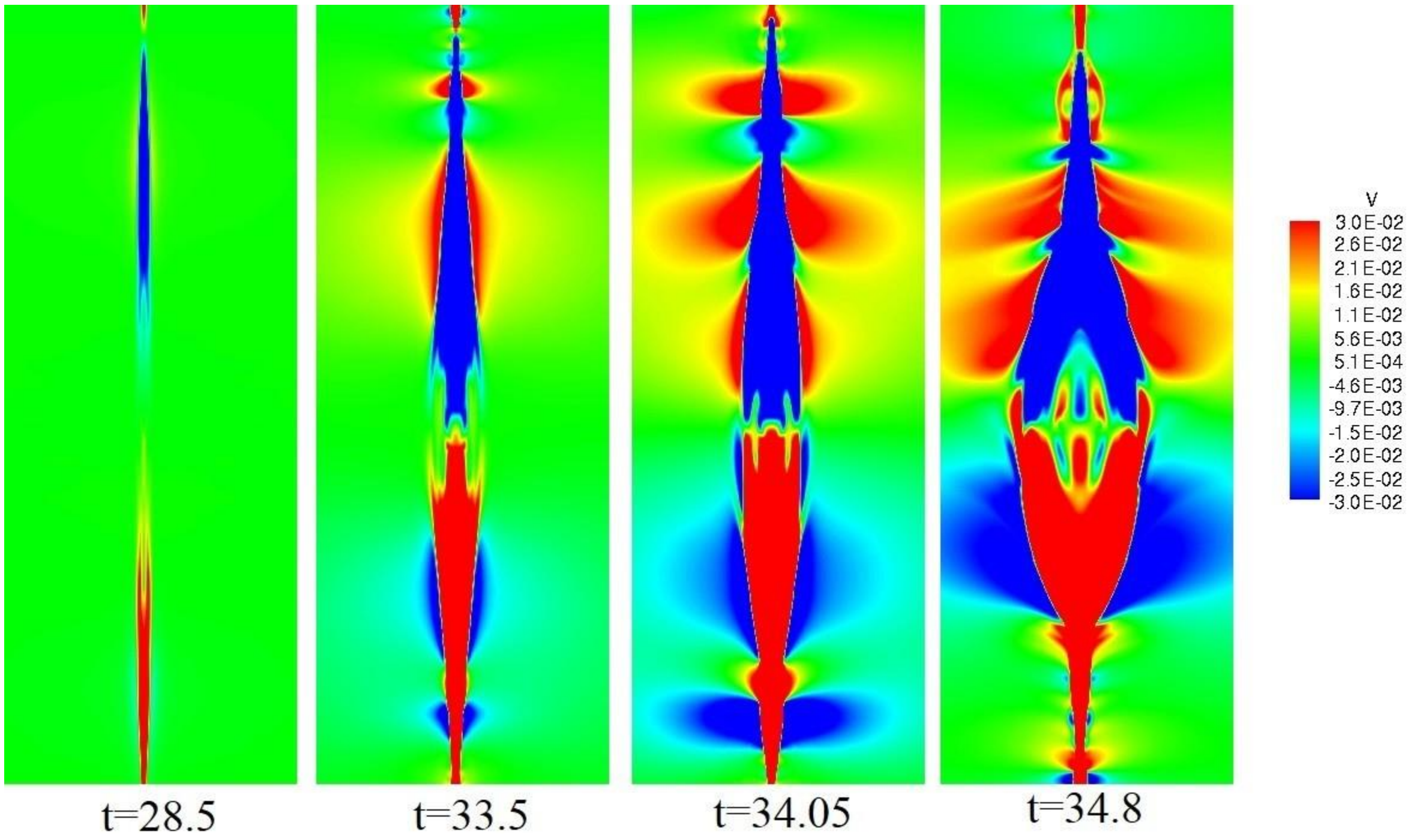

**Figure 20. Contour plots of $\upsilon$ for unsteady magnetic reconnection problem with ideal tearing mode**

# Appendix: Eigensystem of GLM-MHD

The 1D GLM-MHD system of equations has nine eigenvalues. Therefore, the right and left eigenvector matrices expressed in primitive variables in the local reference coordinates have the following forms

$$\mathbb{R}' = [\boldsymbol{r}'_1, \boldsymbol{r}'_2, \boldsymbol{r}'_3, \boldsymbol{r}'_4, \boldsymbol{r}'_5, \boldsymbol{r}'_6, \boldsymbol{r}'_7, \boldsymbol{r}'_8, \boldsymbol{r}'_9]$$
$$\mathbb{L}' = [(\boldsymbol{l}'_1)^T, (\boldsymbol{l}'_2)^T, (\boldsymbol{l}'_3)^T, (\boldsymbol{l}'_4)^T, (\boldsymbol{l}'_5)^T, (\boldsymbol{l}'_6)^T, (\boldsymbol{l}'_7)^T, (\boldsymbol{l}'_8)^T, (\boldsymbol{l}'_9)^T]^T$$

Each of the eigenvalue and its corresponding eigenvector are shown below sequentially,

$$\begin{cases} \text{entropy wave:} \quad \lambda_1 = v_n \\ \boldsymbol{l}'_1 = \left[1 - \dfrac{k}{2a^2}\boldsymbol{v}\cdot\boldsymbol{v}, -\dfrac{kv_n}{a^2}, \dfrac{kv_{t1}}{a^2}, \dfrac{kv_{t2}}{a^2}, 0, \dfrac{kB_{t1}}{a^2}, \dfrac{kB_{t2}}{a^2}, \dfrac{k}{a^2}, 0\right] \\ \boldsymbol{r}'_1 = \left[1, v_n, v_{t1}, v_{t2}, 0,0,0, \dfrac{\boldsymbol{v}\cdot\boldsymbol{v}}{2}, 0\right]^T \end{cases}$$

$$\begin{cases} \text{alfven wave:} \quad \lambda_{2/3} = v_n \pm c_a \\ \boldsymbol{l}'_{2/3} = \left[\dfrac{v_{t1}\beta_{t2} - v_{t2}\beta_{t1}}{\sqrt{2}\rho}, 0, -\dfrac{\beta_{t2}}{\sqrt{2}\rho}, \dfrac{\beta_{t1}}{\sqrt{2}\rho}, 0, \pm\dfrac{\beta_{t2}}{\sqrt{2\rho}}, \mp\dfrac{\beta_{t1}}{\sqrt{2\rho}}, 0,0\right] \\ \boldsymbol{r}'_{2/3} = \left[0,0, -\dfrac{\rho\beta_{t2}}{\sqrt{2}}, \dfrac{\rho\beta_{t1}}{\sqrt{2}}, 0, \pm\sqrt{\dfrac{\rho}{2}}\beta_{t2}, \mp\sqrt{\dfrac{\rho}{2}}\beta_{t1}, \dfrac{\rho(v_{t2}\beta_{t1} - v_{t1}\beta_{t2})}{\sqrt{2}}, 0\right]^T \end{cases}$$

$$\begin{cases} \text{fast magneto-acoustic wave:} \quad \lambda_{4/5} = v_n \pm c_f \\ \boldsymbol{l}'_{4/5} = \Big[\dfrac{1}{2\rho a^2}(\mp\alpha_f c_f v_n \pm \alpha_s c_s sign(B_n)(\beta_{t1}v_{t1} + \beta_{t2}v_{t2}) - \\ \dfrac{1}{2}\alpha_f k\boldsymbol{v}\cdot\boldsymbol{v}), \dfrac{\alpha_f}{2\rho a^2}(\pm c_f + kv_n), \dfrac{1}{2\rho a^2}(\mp\alpha_s c_s\beta_{t1} sign(B_n) + \alpha_f k v_{t1}), \\ \dfrac{1}{2\rho a^2}(\mp\alpha_s c_s\beta_{t2} sign(B_n) + \alpha_f k v_{t2}), 0, \dfrac{\alpha_s\beta_{t1}}{2\sqrt{\rho}a} + \dfrac{\alpha_f k\beta_{t1}}{2\rho a^2}, \\ \dfrac{\alpha_s\beta_{t2}}{2\sqrt{\rho}a} + \dfrac{\alpha_f k\beta_{t2}}{2\rho a^2}, -\dfrac{k\alpha_f}{2\rho a^2}, 0\Big] \end{cases}$$

$$\begin{cases} \boldsymbol{r}'_{4/5} = \big[\rho\alpha_f, \rho\alpha_f(v_n \pm c_f), v_{t1}\rho\alpha_f \mp \alpha_s c_s\beta_{t1} sign(B_n)\rho, \\ v_{t2}\rho\alpha_f \mp \alpha_s c_s\beta_{t2} sign(B_n)\rho, 0, \alpha_s\sqrt{\rho}a\beta_{t1}, \alpha_s\sqrt{\rho}a\beta_{t2}, \\ \dfrac{1}{2}\rho\alpha_f\boldsymbol{v}\cdot\boldsymbol{v} \pm \alpha_s c_f\rho v_n \mp \alpha_s c_s\beta_{t1} sign(B_n)\rho v_{t1} \\ \mp\alpha_s c_s\beta_{t2} sign(B_n)\rho v_{t2} + \alpha_s\sqrt{\rho}a\beta_{t1}B_{t1} + \alpha_s\sqrt{\rho}a\beta_{t2}B_{t2} - \dfrac{\alpha_f\gamma p}{k}, 0\Big]^T \end{cases}$$

$$\begin{cases} \text{slow magneto-acoustic wave:} \quad \lambda_{6/7} = v_n \pm c_s \\ \boldsymbol{l}'_{6/7} = \Big[\frac{1}{2\rho a^2}(\mp\alpha_s c_s v_n \mp \alpha_f c_f sign(B_n)(\beta_{t1}v_{t1} + \beta_{t2}v_{t2}) - \\ \frac{1}{2}\alpha_s k\boldsymbol{v}\cdot\boldsymbol{v}), \frac{\alpha_s}{2\rho a^2}(\pm c_s + kv_n), \frac{1}{2\rho a^2}(\pm\alpha_f c_f \beta_{t1} sign(B_n) + \alpha_s k v_{t1}), \\ \frac{1}{2\rho a^2}(\pm\alpha_f c_f \beta_{t2} sign(B_n) + \alpha_s k v_{t2}), 0, -\frac{\alpha_f \beta_{t1}}{2\sqrt{\rho}a} + \frac{\alpha_s k \beta_{t1}}{2\rho a^2}, \\ -\frac{\alpha_f \beta_{t2}}{2\sqrt{\rho}a} + \frac{\alpha_s k \beta_{t2}}{2\rho a^2}, -\frac{k\alpha_s}{2\rho a^2}, 0\Big] \end{cases}$$

$$\begin{cases} \boldsymbol{r}'_{6/7} = \big[\rho\alpha_s, \rho\alpha_s(v_n \pm c_s), v_{t1}\rho\alpha_s \pm \alpha_f c_f \beta_{t1} sign(B_n)\rho, \\ v_{t2}\rho\alpha_s \pm \alpha_f c_f \beta_{t2} sign(B_n)\rho, 0, -\alpha_f\sqrt{\rho}a\beta_{t1}, -\alpha_f\sqrt{\rho}a\beta_{t2}, \\ \frac{1}{2}\rho\alpha_s \boldsymbol{v}\cdot\boldsymbol{v} \pm \alpha_s c_s \rho v_n \pm \alpha_f c_f \beta_{t1} sign(B_n)\rho v_{t1} \\ \pm\alpha_f c_f \beta_{t2} sign(B_n)\rho v_{t2} - \alpha_f\sqrt{\rho}a\beta_{t1}B_{t1} - \alpha_f\sqrt{\rho}a\beta_{t2}B_{t2} - \frac{\alpha_s \gamma p}{k}, 0\big]^T \end{cases}$$

$$\begin{cases} \nabla\cdot\boldsymbol{B} \text{ wave:} \quad \lambda_{8/9} = \pm c_h \\ \boldsymbol{l}'_{8/9} = \left[0,0,0,0,\frac{1}{2},0,0,0,\pm\frac{1}{2c_h}\right] \\ \boldsymbol{r}'_{8/9} = [0,0,0,0,1,0,0,0,\pm c_h]^T \end{cases}$$

$$\begin{cases} \text{the speed of sound} \quad a = \sqrt{\frac{\gamma p}{\rho}} \\ \text{alfven speed} \quad c_a = \frac{B_n}{\sqrt{\mu_0 \rho}} \\ \text{fast/slow magneto-acoustic speed} \\ c_{f/s} = \sqrt{\frac{1}{2}\left(\frac{\gamma p + \boldsymbol{B}\cdot\boldsymbol{B}}{\rho} \pm \sqrt{\left(\frac{\gamma p + \boldsymbol{B}\cdot\boldsymbol{B}}{\rho}\right)^2 - 4\frac{\gamma p B_n^2}{\rho^2}}\right)} \\ k = 1 - \gamma \end{cases}$$

Finally, all eigenvectors are scaled by parameters given by Roe & Balsara[17]

$$\alpha_f^2 = \frac{a^2 - c_s^2}{c_f^2 - c_s^2} \qquad \alpha_s^2 = \frac{c_f^2 - a^2}{c_f^2 - c_s^2}$$

$$\beta_{t1} = \frac{B_{t1}}{\sqrt{B_{t1}^2 + B_{t2}^2}} \quad \beta_{t2} = \frac{B_{t2}}{\sqrt{B_{t1}^2 + B_{t2}^2}}$$

$$\text{when} \quad B_{t1}^2 + B_{t2}^2 = 0, \quad \beta_{t1} = \beta_{t2} = \frac{1}{\sqrt{2}}$$